\documentclass[twocolumn]{aastex62}

\begin{document}

\shorttitle{X-ray Obscuration in 3C 223}
\shortauthors{LaMassa et al.}

\title{The Complex X-ray Obscuration Environment in the Radio Loud Type 2 Quasar 3C 223}

\correspondingauthor{Stephanie LaMassa}
\email{slamassa@stsci.edu}

\author[0000-0002-5907-3330]{Stephanie M. LaMassa}
\affiliation{Space Telescope Science Institute,
3700 San Martin Drive,
Baltimore, MD, 21218, USA}

\author{Tahir Yaqoob}
\affiliation{CRESST,
University of Maryland Baltimore County,
1000 Hilltop Circle, Baltimore, MD, 21250, USA}
\affiliation{NASA/Goddard Spaceflight Center,
Mail Code 662,
Greenbelt 20771, USA}

\author[0000-0001-5737-5055]{Panayiotis Tzanavaris}
\affiliation{University of Maryland Baltimore County,
1000 Hilltop Circle, Baltimore, MD, 21250, USA}
\affiliation{Laboratory for X-ray Astrophysics,
NASA/Goddard Spaceflight Center,
Mail Code 662,
Greenbelt 20771, USA}
\affiliation{CRESST,
University of Maryland Baltimore County,
1000 Hilltop Circle, Baltimore, MD, 21250, USA}

\author[0000-0003-3105-2615]{Poshak Gandhi}
\affiliation{School of Physics \& Astronomy,
  University of Southampton,
  Highfield, Southampton SO17 1BJ, UK}
\affiliation{Inter-University Centre for Astronomy \& Astrophysics,
  Post Bag 4,
  Ganeshkhind, Pune 411007, India}

\author[0000-0001-6670-6370]{Timothy Heckman}
\affiliation{Center for Astrophysical Sciences,
  Department of Physics \& Astronomy,
  The Johns Hopkins University,
  Baltimore, MD 21218, USA}

\author[0000-0002-5328-9827]{George Lansbury}
\affiliation{European Southern Observatory,
  Karl-Schwarzschild str. 2, D-85748
  Garching bei München, Germany}

\author{Aneta Siemiginowska}
\affiliation{Center for Astrophysics,
  Harvard and Smithsonian,
  60 Garden St.,
  Cambridge, MA 02138, USA}

\begin{abstract}
  3C 223 is a radio loud, Type 2 quasar at $z=0.1365$ with an intriguing {\it XMM-Newton} spectrum that implicated it as a rare, Compton-thick ($N_{\rm H} \gtrsim 1.25 \times 10^{24}$ cm$^{-2}$) active galactic nucleus (AGN). We obtained contemporaneous {\it XMM-Newton} and {\it NuSTAR} spectra to fit the broad-band X-ray spectrum with the physically-motivated \textsc{MYTorus} and \textsc{borus02} models. We confirm earlier results that the obscuring gas is patchy with both high (though not Compton-thick) levels of obscuration ($N_{\rm H} > 10^{23}$ cm$^{-2}$) and gas clouds with column densities up to an order of magnitude lower. The spectral fitting results indicate additional physical processes beyond those modeled in the spectral grids of \textsc{MYTorus} and \textsc{borus02} impact the emergent spectrum: the Compton-scattering region may be extended beyond the putative torus; a ring of heavy Compton-thick material blocks most X-ray emission along the line of sight; or the radio jet is beamed, boosting the production of Fe K$\alpha$ line photons in the global medium compared with what is observed along the line of sight. We revisit a recent claim that no radio loud Compton-thick AGN have yet been conclusively shown to exist, finding three reported cases of radio loud AGN with global average (but not line-of-sight) column densities that are Compton-thick. Now that it is possible to separately determine line-of-sight and global column densities, inhomogeneity in the obscuring medium has consequences for how we interpet the spectrum and classify an AGN as ``Compton-thick.''
\end{abstract}


\section{Introduction}

The hidden side of supermassive black hole accretion has long captivated the astronomy community. The observational properties of active galactic nuclei (AGN) are diverse \citep{hickoxalexander}, but can trace their origins to accretion onto a black hole that ionizes gas both near and far (in the Broad Line Region and Narrow Line region, respectively). A corona of hot gas above and below the accretion disk is thought to inverse Compton scatter optical and ultraviolet photons from the disk to X-ray energies \citep{haardtmaraschi1991}. In some AGN, a radio jet is launched from either the inner edges of the accretion disk \citep{blandfordpayne} or via a mechanism involving magnetic fields tapping into the black hole spin energy \citep{blandfordznajek}. According to AGN unified models, a ``torus'' of gas and dust surrounds the accretion disk, absorbing photons along the equatorial plane but allowing radiation from the accretion disk to escape along the polar axis \citep{antonucci,urrypadovani,netzer}. Dust in the torus absorbs the optical to ultraviolet light emanating from the accretion disk while gas, which can exist within the dust sublimation zone, absorbs and scatters X-ray photons produced by the corona. When the column density of this obscuring gas reaches levels where it becomes optically thick to Compton scattering ($N_{\rm H} \geq$ 1/(1.2 $\times \sigma_{T}$) $\simeq 1.25 \times 10^{24}$ cm$^{-2}$)\footnote{$\sigma_{T}$ is the Thompson cross-section and the factor of 1.2 accounts for the contributions of electrons from both Hydrogen and Helium.}, the AGN is defined as ``Compton-thick.''

Compton-thick AGN are an interesting population that, though difficult to detect, nevertheless leave their imprint on the cosmic energy budget and cannot be dismissed as mere curiosities. Studies of the cosmic X-ray background reveal an unresolved component: after integrating the emission of resolved X-ray sources, excess emission remains \citep[e.g.,][]{worsley}. A sizeable portion of Compton-thick AGN have been invoked to account for this excess \citep{ueda2003}. The exact fraction of accretion enshrouded by Compton-thick material is still a matter of debate as different model assumptions about the X-ray spectra of AGN and observational biases of X-ray surveys can lead to different results. Currently the Compton-thick AGN fraction has been quoted to be anywhere from $\sim9 - 50$\% \citep{gilli,treister,akylas,ueda,ricci2015,ananna} though these studies do not incorporate models where the line-of-sight and global column densities can differ significantly as reported in detailed studies of individual AGN \citep[e.g.,][]{yaqoob2015,lamassa2016,lamassa2017,balokovic,tzanavaris2019,tzanavaris2021,zhao}.

Compton-thick AGN may represent a transient phase in supermassive black hole evolution. For instance, the major merger paradigm of black hole growth predicts a phase of rapid black hole growth nestled within high-to-Compton-thick levels of obscuration, followed by AGN outflows that evacuate the cocoon of absorbing material \citep{sanders,hopkins2006,hopkins2008}. A better understanding of the Compton-thick population, and the complex distribution of obscuring gas, can then shed light about a key phase of galaxy and black hole co-evolution.

Radio jets launched from an AGN also play a role in shaping the environment in which an AGN lives, both on galactic and extragalactic scales \citep[see][and references therein]{fabian2012}. Recently, \citet{ursini} claimed that no bona-fide Compton-thick radio loud AGN have demonstrably been identified. Though a handful of radio loud Compton-thick {\it candidates} have been proposed \citep{panessa2016}, the column densities for these sources were estimated from X-ray spectral fits below 10 keV, often using simplified spectral models. Simple absorbed power law models that assume a screen of extinction often fail to capture the inherent complexity of AGN spectra and can under-estimate the obscuring column density when the spectrum is dominated by scattered AGN emission that leaks through the torus \citep{turner1997,turner2009,winter2009,lamassa2009,lamassa2011}. Other clues can point to heavy levels of obscuration, including a depressed observed 2-10 keV X-ray flux when normalized to the intrinsic AGN luminosity \citep{bassani1999,heckman2005,panessa2006,lamassa2009,lamassa2011,jia}, and a large Fe K$\alpha$ equivalent width (EW $\gtrsim$ 1 keV) due to the supression of the absorbed continuum against which the Fe K$\alpha$ line (formed via fluoresence within the obscuring matter) is measured \citep{krolik,ghisellini,matt}, though weaker Fe K$\alpha$ lines are sometimes observed in Compton-thick AGN \citep[see][]{gandhi2017, boorman2018}.

Even when more complex models are invoked that self-consistently account for the effects of Compton-scattering and therefore measure the column density more accurately, spectra below 10 keV only tell part of the story. Spectra above 10 keV are necessary to observe the Compton hump, thereby constraining the spectral fit. Such broad band spectral fitting (0.5-50 keV) sometimes paints very different pictures about the column density and geometry of the X-ray obscuring medium than when only data below 10 keV are fitted \citep[see, e.g.,][]{lamassa2019}. Indeed, \citet{ursini} point out that when including $>$10 keV {\it NuSTAR} data in the spectral fitting of three radio loud Compton-thick candidates, none of them had measured column densities within the Compton-thick regime. 

Here we explore the X-ray properties of 3C 223, a radio loud, Type 2 (i.e., optically obscured) quasar at $z=0.1365$. It is a Fanaroff-Riley class II \citep[FR II, radio lobe dominated;][]{fanaroffriley}, High Excitation Radio Galaxy \citep[strong narrow optical emission lines;][]{hinelongair} with a projected linear size larger than 1 Mpc where the age of the radio source is estimated to be $72 \pm 4$ Myr \citep{orru}. \citet{reyes} selected this source as a Type 2 quasar on the basis of its high [\ion{O}{3}] luminosity. \citet{lamassa2014} presented 3C 223\footnote{3C 223 was referred to by its SDSS identifier, J093952.74+355358.0, in \citet{reyes}, \citet{jia}, and \citet{lamassa2014}.} as a Compton-thick candidate, where its unusual X-ray spectral properties observed by {\it XMM-Newton} were highlighted. This source boasted a large Fe K$\alpha$ EW \citep[$\sim$500 eV,][]{jia} while lacking an accompanying spectral curvature between 2 - 6 keV associated with Compton-thick reprocessing. \citet{lamassa2014} posited that the spectral features were a clue that the X-ray reprocessing medium was non-uniform, with a global average column density that is Compton-thin (consistent with the observed lack of spectral curvature between 2 - 6 keV), but a line-of-sight column density that is Compton-thick but with a low covering factor. They proposed that a ``ring'' of Compton-thick obscuration could suppress the continuum just enough around the Fe K$\alpha$ line to boost the EW, but not impose the spectral curvature expected from a medium that is globally Compton-thick. They found a good fit to the {\it XMM-Newton} spectrum of 3C 223 using the physically motivated \textsc{MYTorus} model \citep{mytorus} in a configuration where they decoupled the line-of-sight column density ($N_{\rm H,los}$) from the global average column density ($N_{\rm H,global}$). They measured a Compton-thick line-of-sight column density ($N_{\rm H,los} > 1.7 \times10^{24}$ cm$^{-2}$) and a Compton-thin global average column density ($N_{\rm H,global} = 1.4 \times10^{23}$ cm$^{-2}$), supporting the hypothesis that a ring of Compton-thick gas clouds were embedded in a global Compton-thin medium.

Could 3C 223 then be an exception to the trend reported by \citet{ursini}? We obtained contemporaneous {\it NuSTAR} and {\it XMM-Newton} data to find out. In Section \ref{spec_fit}, we describe the results from fitting the {\it XMM-Newton} and {\it NuSTAR} spectra with the \textsc{MYTorus} and \textsc{borus02} models, with a summary of rejected model fits described in the Appendix. In Section \ref{obsc_med}, we explore possible physical scenarios that can explain the X-ray fitting results. From the absorption corrected X-ray luminosity and [\ion{O}{3}] luminosity, we estimate the bolometric luminosity and Eddington ratio of 3C 223 (Section \ref{edd}) and compare that with amount of energy carried by the radio jet in Section \ref{radio}. In Section \ref{qso_comp}, we compare 3C 223 with other [\ion{O}{3}]-defined Type 2 quasars observed by {\it NuSTAR}. Finally, we revisit the question of whether there is a lack of radio loud Compton-thick AGN in Section \ref{cthick_dearth}. Throughout the paper, we assume a cosmology where H$_{\rm 0}$ = 67.7 km/s/Mpc and $\Omega_{m}$ = 0.307 \citep{planck2015}. 

\section{X-ray Observations and Data Reduction}
        {\it XMM-Newton} observed 3C 223 in 2001 October for 35 ks (PI: Birkinshaw, ObsID: 0021740101) and in 2019 November for 48 ks (PI: LaMassa; ObsID: 0852580101). {\it XMM-Newton} is sensitive to X-ray light with energies between 0.5 - 10 keV and has 3 co-aligned detectors: PN, MOS1, and MOS2. We processed the data from all three detectors with the {\it XMM-Newton} Science Analysis System (\textsc{SAS}) v.1.3. We applied standard filtering to the data and assigned good time intervals by inspecting the light curves and using a count rate threshold to remove periods of flaring from the particle background. As shown in Table \ref{Observing_Log}, this filtering leaves us with a net exposure of about 15-19 ks and 17-23 ks for the observations from 2001 and 2019 respectively.
        
        We extracted spectra from the PN, MOS1, and MOS2 detectors using a circular aperture with a radius of 30$^{\prime\prime}$ ($\sim$75 kpc) centered on the X-ray source. The background counts for all three detectors were extracted from a source-free circular region with a 75$^{\prime\prime}$ radius near the AGN. The net counts summed among the three detectors for the observations from 2001 and 2019 are 1526.6 and 1391.5, respectively.  \citet{croston} identified X-ray emission from the radio lobes of 3C 223 in the 2001 {\it XMM-Newton} observation of this source which they attributed to inverse Compton scattering from cosmic microwave background photons within the lobes. Since the purpose of our investigation is to understand the circumnuclear obscuration environment, we took care to extract the source and background spectra from regions that did not overlap the X-ray lobes reported by \citet{croston}.
        
        3C 223 was observed with {\it NuSTAR} in 2019 November (PI: LaMassa, ObsID: 60501020002) for 48.5 ks. {\it NuSTAR} is the first hard X-ray ($>$10 keV) focusing X-ray observatory and has two detectors in the focal plane (FPMA and FPMB) which are sensitive to energies between 3 - 79 keV \citep{harrison}. We processed the data with the \textsc{nupipeline} package of the {\it NuSTAR} Data Analysis Software (\textsc{NuSTARDAS}) v.2.0.0 using CALDB v.20210701. We extracted spectra separately from both modules, using a circular source region with a 45$^{\prime\prime}$ ($\sim$110 kpc) radius and a circular background region of radius 90$^{\prime\prime}$ chosen to be on the same quadrant of the detector as the source to best sample the local background. Between the two modules, we detected 1070.6 net X-ray photons from the {\it NuSTAR} observation of 3C 223. 
        
        We grouped the spectra using \textsc{ftgrouppha} to have a minimum signal-to-noise ratio (S/N) in the background-subtracted spectra of S/N = 3 for the {\it XMM-Newton} spectra and S/N = 2 for the {\it NuSTAR} spectra. This binning allows for adequate resolution around the Fe K$\alpha$ line in the {\it XMM-Newton} spectra, while the lower S/N threshold for the {\it NuSTAR} spectra allows for a spectral bin between 20 - 30 keV.

\begin{deluxetable*}{lllll}
\tablecaption{\label{Observing_Log}X-ray Observing Log of 3C 223}
\tablehead{\colhead{Observatory} & \colhead{Observation Date} &  \colhead{Observation ID} &
  \colhead{Exposure Time (ks)\tablenotemark{a}} & \colhead{Net Counts} \\
}
\startdata
{\it XMM-Newton} & 2001-10-27 & 0021740101 & 34.9           & 1526.6 \\
PN/MOS1/MOS2     &            &            & 15.7/18.2/19.4 & 890.6/328.5/307.5 \\ 
{\it XMM-Newton} & 2019-11-09 & 0852580101 & 48.0           & 1391.5 \\
PN/MOS1/MOS2     &            &            & 17.2/23.1/23.5 & 761.5/316.7/313.3 \\
{\it NuSTAR}     & 2019-11-10 & 6050102000 & 48.5           & 1070.6 \\
FPMA/FPMB       &            &            & 48.5/48.2      & 517.4/553.2 \\
\enddata
\tablenotetext{a}{The top row reports the total duration of the X-ray observations. The second row reports the net exposure time after filtering the data and applying the good time intervals (GTI). Since the data from each detector were reduced independently, there are different GTIs for each dataset, resulting in different next exposure times.}
\end{deluxetable*}

\section{X-ray Spectral Fitting}
We used \textsc{XSpec} v.12.11.1 \citep{arnaud} to fit the {\it XMM-Newton} spectra from 2001 and 2019 simultaneously with the {\it NuSTAR} spectrum from 2019, using $\chi^2$ as the fit statistic. All reported errors represent the 90\% confidence interval ($\Delta \chi^2$=2.706). As we discuss in detail in Appendix \ref{xray_var}, we find no evidence for spectral variability among the X-ray observations of 3C 223.

\subsection{Self-Consistent Treatment of X-ray Reprocessing}

\subsubsection{MYTorus Model}
We fitted the spectra of 3C 223 with \textsc{MYTorus}, a physically motivated X-ray model that self-consisently treats the reprocessing of X-ray emission in obscured AGN \citep[$N_{\rm H} > 10^{22}$ cm$^{-2}$;][]{mytorus}. The \textsc{MYTorus} model is available as \textsc{XSpec} table models that are derived from Monte Carlo simulations that calculate the effects of the transmitted continuum, Compton scattered component, and Fe K$\alpha$ and Fe K$\beta$ fluorescent line emission over a range of input parameters to produce observed X-ray spectra. The elemental abundances are the solar values from \citet{anders} with photoelectric absorption cross-sections from \citet{verner}. 

The default configuration of the \textsc{MYTorus} model assumes a geometry where the X-ray reprocessor is ``doughnut-shaped'' (azimuthally symmetric), and has a fixed covering factor of 0.5, corresponding to an opening angle of 60$^{\circ}$. In this default configuration, any inclination angle greater than 60$^{\circ}$ intersects the torus while any smaller inclination angle corresponds to a face-on orientation that allows an unimpeded view of the X-ray emitting region. Modeling the intrinsic AGN continuum as a powerlaw, the allowed ranges of the spectral slope within the \textsc{MYTorus} model are $1.4 \leq \Gamma \leq 2.6$. The allowed equivalent hydrogen column densities range from 10$^{22}$ cm$^{-2}$ - 10$^{25}$ cm$^{-2}$, and in this default, ``coupled'' configuration, the measured column density reflects the equatorial column density ($N_{\rm H,equatorial}$), i.e., the amount of Hydrogen measured through the diameter of the torus cross-section.

However, the X-ray reprocessor need not be uniform, and mounting evidence indicates that the X-ray reprocessor is not always homogeneous, with significantly different global average column densities from those measured along the line of sight \citep{yaqoob2015,lamassa2016,lamassa2017,balokovic,tzanavaris2019,tzanavaris2021,zhao}. Eclipsing events that have been observed in the X-ray spectra of some AGN offer clues that discrete clouds of gas can transit into and out of the line of sight, manifesting significant changes in the observed column density on the timescale of days to years \citep{mckernan,risaliti2002,risaliti2005,rivers2015,marinucci2016,ricci2016}. The decoupled \textsc{MYTorus} model configuration offers the flexibility to disentangle the global column density ($N_{\rm H,global}$) from the line-of-sight column density ($N_{\rm H,los}$) \citep[see][]{yaqoob2012,lamassa2014,tzanavaris2019}. The geometry or covering fraction of the X-ray reprocessor can not necessarily be constrained by the decoupled \textsc{MYTorus} model configuration, but measurements of the independent column densities indicate whether the obscuring medium is patchy or uniform.

For both the coupled and decoupled \textsc{MYTorus} configurations, the high-level model set-up in \textsc{XSpec} is the same:
\begin{eqnarray}
  \begin{array}{ll}
  \mathrm{model} = & const_1 \times phabs \times \nonumber \\
  & (zpow_1 \times \mathrm{etable\{mytorus\_Ezero\_v00.fits\}} + \nonumber \\
  & const_2 \times \mathrm{atable\{mytorus\_scatteredH200\_v00.fits\}} + \nonumber \\
  & const_3 \times  \nonumber \\
  & \mathrm{atable\{mytl\_V000010nEp000H200\_v00.fits\}})

  \end{array}
\end{eqnarray}

\noindent where \textsc{zpow\_1} models the intrinsic AGN powerlaw emission; \textsc{mytorus\_Ezero\_v00.fits}, \\ \textsc{mytorus\_scatteredH200\_v00.fits}, and \\ \textsc{mytl\_V000010nEp000H200\_v00.fits} model the attenuation of the transmitted AGN continuum, Compton scattered emission, and fluorescent line emission, respectively; \textsc{phabs} is frozen to the Galactic column density ($N_{\rm H,Galactic} = 1.46 \times 10^{20}$ cm$^{-2}$);\footnote{Galactic column density derived from Colden which is hosted by the {\it Chandra} X-ray Observatory science center: https://cxc.harvard.edu/toolkit/colden.jsp} and \textsc{const\_1}, \textsc{const\_2}, and \textsc{const\_3} are constants to model the cross-instrument normalization, relative normalization between the transmitted and Compton scattered components ($A_{\rm S}$), and relative normalization between the transmitted and fluorescent line emission component ($A_{\rm L}$, which is always linked to $A_{\rm S}$), respectively. The column density and inclination angle associated with the Compton scattered and fluorescent line emission components are always tied to the same value. 

For the coupled \textsc{MYTorus} configuration, all torus components (i.e., $N_{\rm H,equatorial}$ and inclination angle) are linked between the transmitted, and Compton scattered/fluorescent line models. In the decoupled configuration, the column densities are fitted independently between the transmitted component and the Compton scattered/fluorescent line emission. The inclination angle of the X-ray reprocessor associated with the transmitted component is frozen to 90$^{\circ}$ so that the measured column density is that along the line of sight ($N_{\rm H,los}$). The inclination angle of the X-ray reprocessor associated with the Compton scattered and fluorescent line emission components is frozen to 0$^{\circ}$ to mimic face-on reflection via scattering off the back-side of the torus without further interaction with intervening matter. Here, the fitted column density reflects the global average column density ($N_{\rm H,global}$).

\subsubsection{BORUS02 Model}

The \textsc{borus02} model \citep{balokovic} allows the global average column density to be fitted independently from the line-of-sight column density, similar to the \textsc{MYTorus} model in decoupled mode. Unlike the \textsc{MYTorus} model, this independence in fitted column densities is achieved by adding an additional cutoff powerlaw model component (\textsc{cutoffpl}) to the overall model that is distinct from the \textsc{BORUS} spectral templates calculated via radiative transfer codes from which the \textsc{borus02} XSpec table model is defined. Self-consistency between the absorbed power law component and \textsc{borus02} model is assumed by linking the parameters of the powerlaw model to the \textsc{borus02} model. 

\begin{eqnarray}
  \begin{array}{ll}
  \mathrm{model} = & const_1 \times phabs \times \nonumber \\
  & (\mathrm{atable\{borus02\_v170323a.fits\}} + \nonumber \\
  & zphabs \times cabs \times \mathrm{cutoffpl})
  \end{array}
\end{eqnarray}

Here, the first \textsc{phabs} component is frozen to the Galactic column density (as above) while the second \textsc{phabs} component is linked to the \textsc{cabs} column density, which is an absorption model that accounts for Compton-scattering. 

\textsc{borus02} also fits the covering factor of the torus ($CF_{\rm tor}$ = cos($\theta_{\rm tor}$), where $\theta_{\rm tor}$ is the torus opening angle), which is the fraction of the sky obscured by the torus as seen by the X-ray source. The covering factor ranges from 10\% to 100\% (i.e., completely covered). However, as noted in \citet{lamassa2019}, it us unclear how this covering factor relates to the line-of-sight column density as the latter is parameterized by an additive model component that is independent from the radiative transfer calculations used to derive the \textsc{borus02} spectral template grid in which $CF_{\rm tor}$ is measuresed. An advantage of the \textsc{borus02} model is that the iron abundance ($A_{\rm Fe}$) can be fitted as a free parameter, allowing us to test whether a super-solar iron abundance can partially explain the enhanced iron emission with respect to the continuum seen in the {\it XMM-Newton} spectra \citep{jia,lamassa2014}. The \textsc{borus02} energy lower energy range terminates at 1 keV, unlike the \textsc{MYTorus} model which is valid to 0.6 keV.

\subsection{Modeling the Soft X-ray Emission: Scattered AGN Light or Jet Emission?}
Excess emission above an absorbed power law (representing the attenuated transmitted AGN continuum) is apparent at soft X-ray energies ($<$2 keV) for many AGN, and can originate from different components. This emission can represent the scattering of the intrinsic AGN continuum off of free electrons in distant, optically thin material \citep{winter2009,turner2009,lamassa2009,lamassa2011}; thermal emission from the host galaxy \citep{turner1997,lamassa2012} or unresolved photoionized emission due to the AGN \citep[e.g.,][]{guianazzi2007}; or non-thermal unresolved jet emission \citep{hardcastle2006,hardcastle2009}. In low-resolution (i.e., non-grating, non-microcalorimeter) X-ray spectra from CCD detectors, thermal emission from star-formation or AGN photoionization can be well described with thermal X-ray models like \textsc{apec}, where high resolution spectroscopy is necessary to definitively pinpoint the physical origins of the soft emission. Though the soft X-ray spectra of radio loud AGN can often be well fitted by either of the non-thermal models, \citet{hardcastle2006,hardcastle2009} argue that the soft X-ray emission in radio loud AGN has its physical origins in an X-ray jet due to the strong correlation between the nuclear radio (jet) emission and the soft X-ray luminosity. 

Both a partial covering and a jet-origin scenario are described by powerlaw models, so we added a powerlaw model to the above model set-ups to accommodate the soft X-ray spectrum below 2 keV. This powerlaw emission is attenuated by $N_{\rm H,Galactic}$ but is unaffected by the circumnuclear absorption that suppresses the AGN emission. 

In the partial covering model, this second power law model describes AGN emission that is scattered into the line of sight. It therefore has the same power law slope ($\Gamma$) and normalization as the primary power law model, where this consistency is enforced by linking these parameters together in the spectral fitting. A constant multiplicative factor was introduced as a free parameter before the second power law model to parameterize the scattering fraction ($f_{\rm scatt}$). To describe soft X-ray emission from a jet, the spectral slope and normalization of the second power law component are fitted independently from the primary power law, and no exta multiplicative factor is included in the fitting. \citet{hardcastle2009} find that the slope of the unabsorbed power law component (which they ascribe to the jet) varies from $1.2 \lesssim \Gamma \lesssim 2.4$, with a median value of $\Gamma$ = 1.6 (calculated from the fits where $\Gamma$ to the unabsorbed power law was a free parameter in the modeling). This median powerlaw slope is somewhat harder than the canonical AGN slope of $ 1.7 < \Gamma < 1.9$ \citep[e.g.,][]{tozzi2006}, but there is a wide range of fitted values for the assumed jet component, so there is not a typical power law slope that cleanly distinguishes AGN coronal emission from jet emission.

Since the \textsc{MYTorus} model is sensitive to lower energies than the \textsc{borus02} model, we use the former model to try to distinguish between a jet or scattered AGN light origin for the soft X-ray emission. With the \textsc{borus02} model, we simply accommodate the excess between 1 - 2 keV with a cutoff powerlaw model modified by a multiplicative factor to parameterize $f_{\rm scatt}$.

\subsection{X-ray Fitting Results}\label{spec_fit}
\begin{figure*}[h]
  \begin{center}
    \includegraphics[scale=0.35]{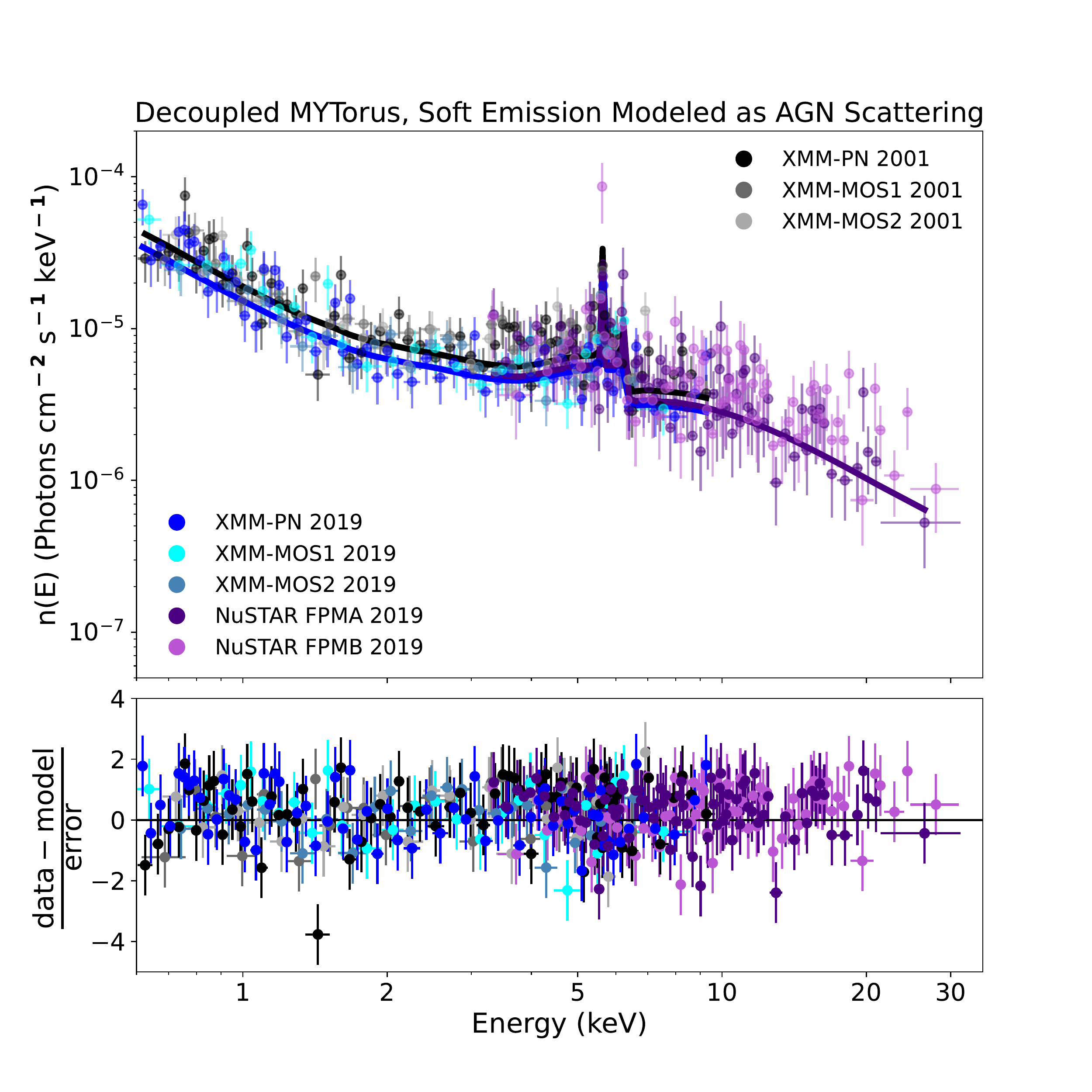}~
    \hspace{-0.3cm}
  \includegraphics[scale=0.47]{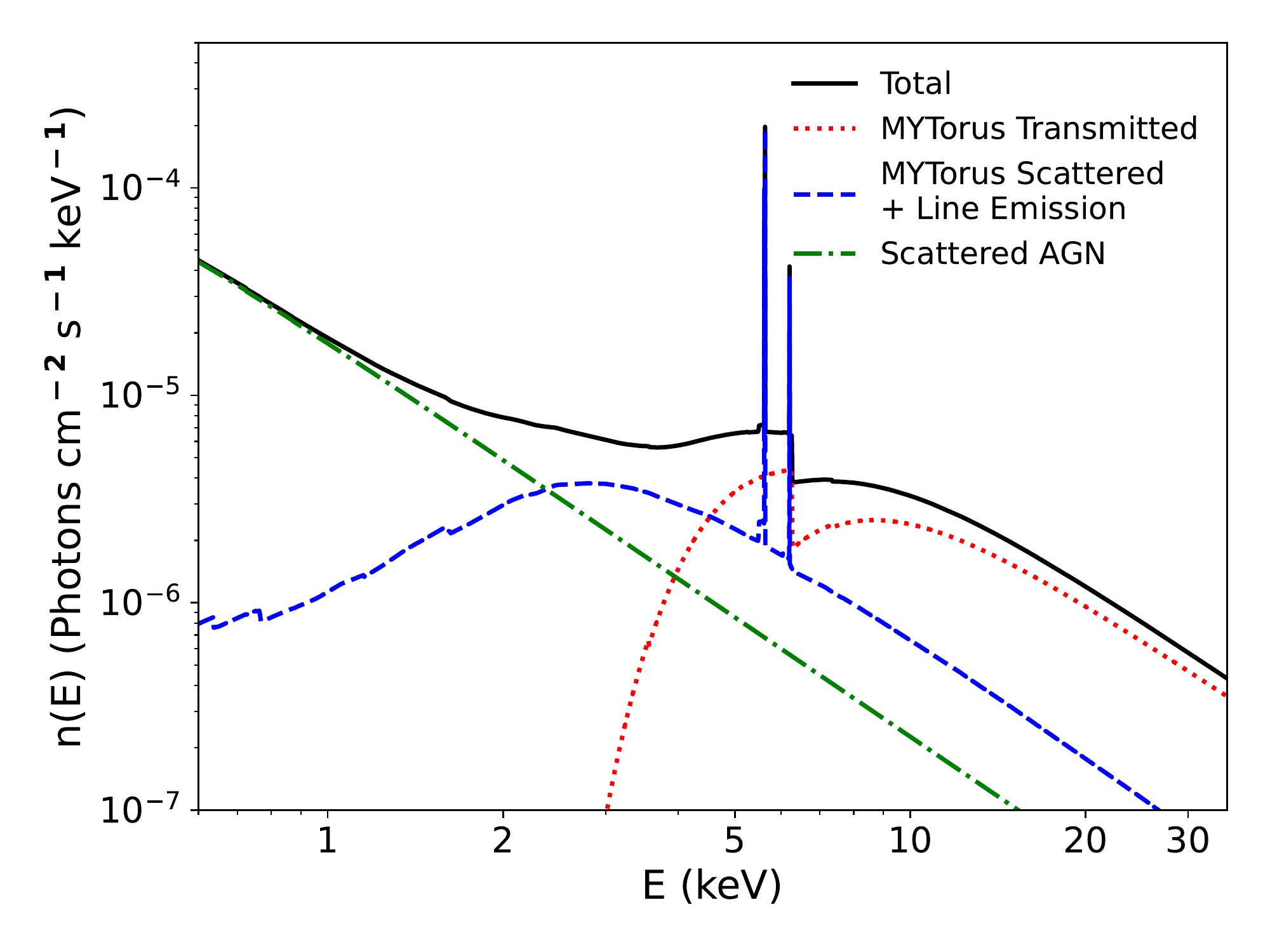}\\
  \includegraphics[scale=0.35]{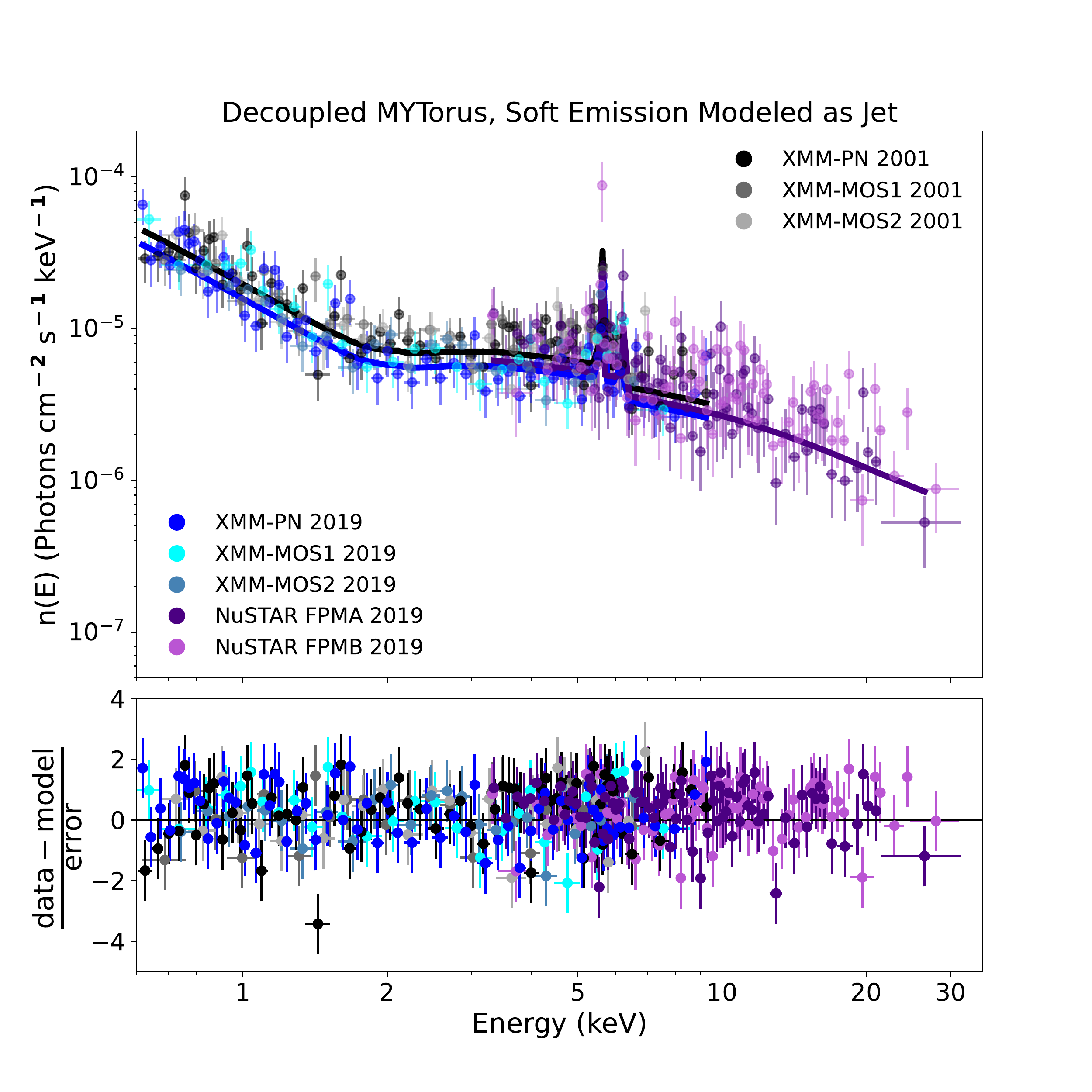}~
  \hspace{-0.3cm}
  \includegraphics[scale=0.47]{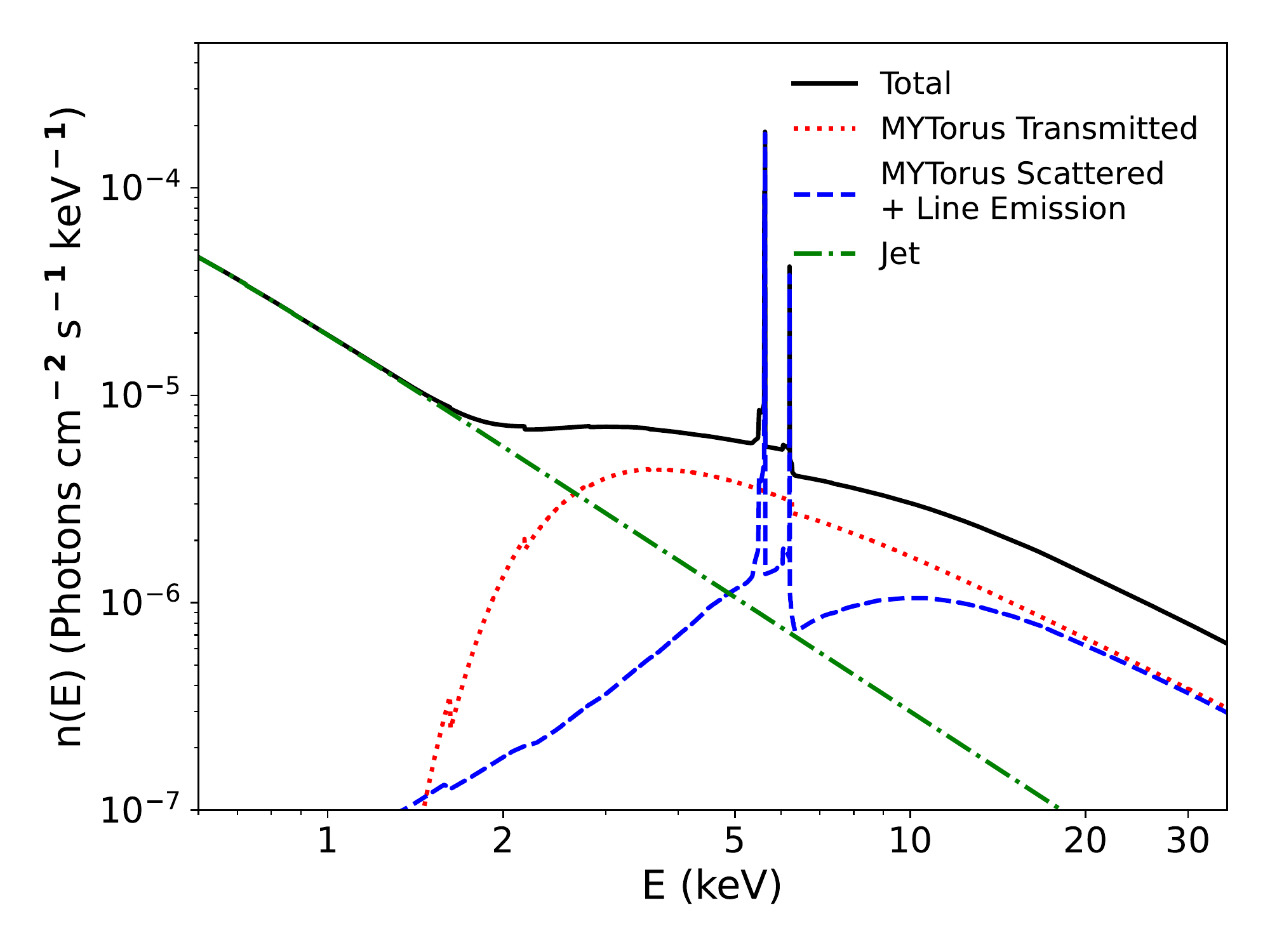}
  \caption{\label{mytor_decoup_freeas_plots} {\it Left}: Decoupled \textsc{MYTorus} fit to the unfolded {\it XMM-Newton} and {\it NuSTAR} spectra of 3C 223, modeling the soft X-ray emission (0.5 - 2 keV) as scattered AGN light (top) and as jet emission (bottom). {\it Right}: Best-fit model, with individual components plotted separately for illustrative purposes. Both models allow the normalization between the transmitted AGN continuum and Compton scattered/fluorescent line emission to be a free parameter. The models provide nearly identical fits to the spectra, but with different intrinsic physical properties. With the scattered AGN model, the X-ray spectrum is reflection dominated between 2 - 5 keV while it is transmission dominated above 2 keV in the jet model.}
  \end{center}
\end{figure*}

\begin{figure*}[h]
  \begin{center}
  \includegraphics[scale=0.35]{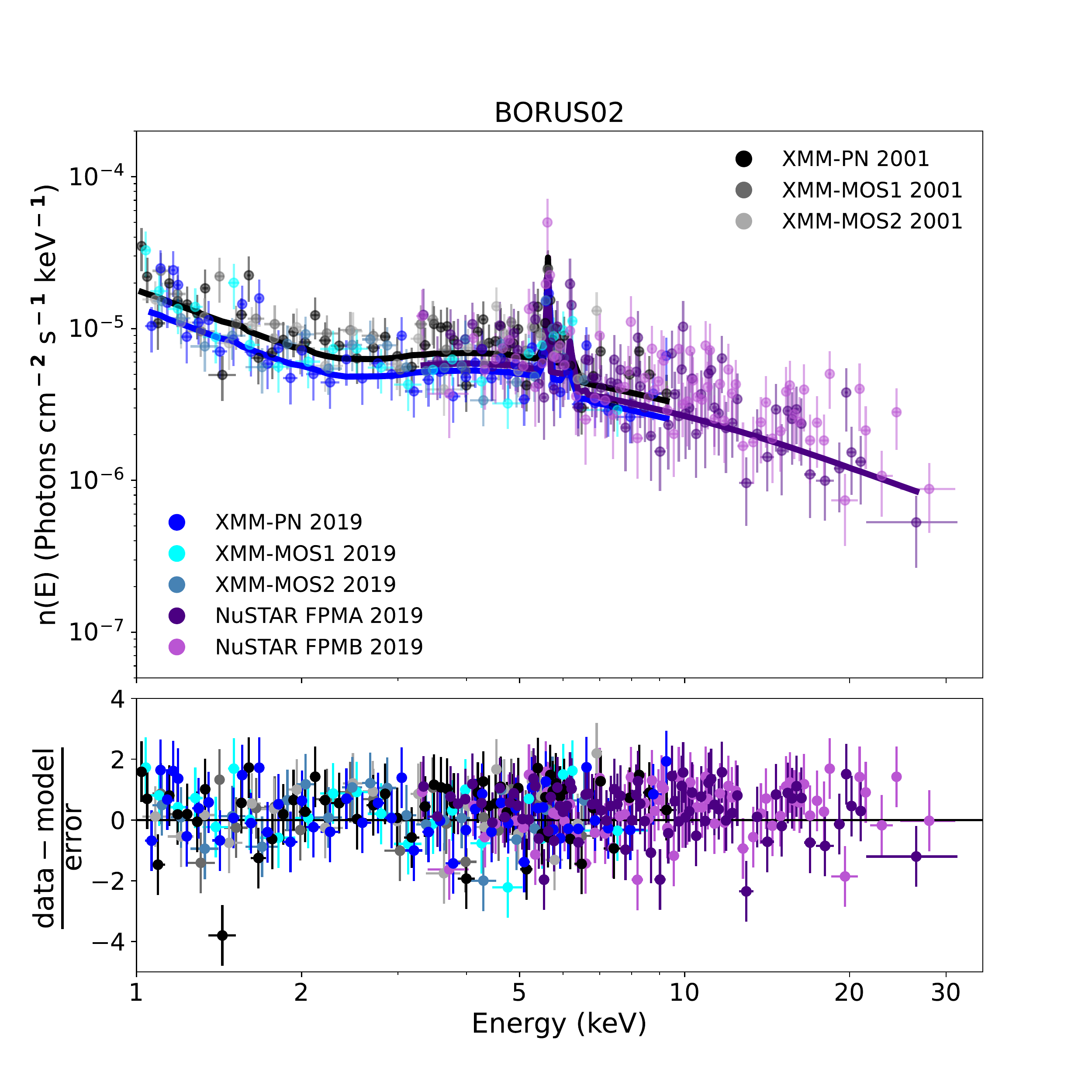}~
  \hspace{-0.3cm}
  \includegraphics[scale=0.47]{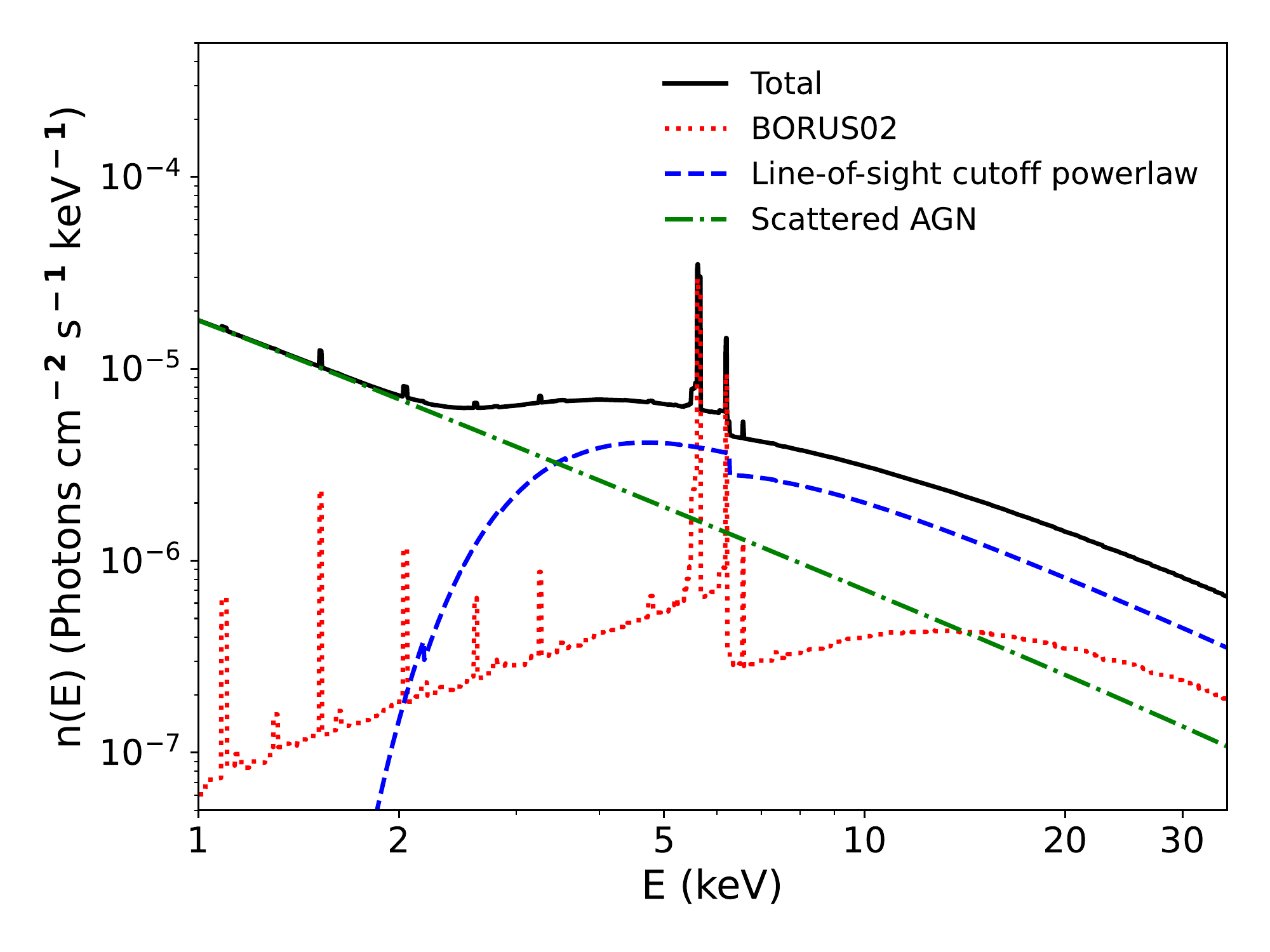}
  \caption{\label{borus02_plots} {\it Left}: The unfolded {\it XMM-Newton} and {\it NuSTAR} spectra fitted with the \textsc{borus02} model, which has a lower energy cutoff of 1 keV that is higher than that of \textsc{MYTorus}. {\it Right}: Best-fit model with individual model components plotted separately. According to this model, the spectrum is transmission dominated above 3 keV, with scattered AGN light being the main contributor to the X-ray spectrum at lower energies (compared with the \textsc{MYTorus} model fits shown in Figure \ref{mytor_decoup_freeas_plots} where the scattered AGN component becomes dominant at energies below 2 keV).}
  \end{center}
\end{figure*}

We found that the X-ray spectra of 3C 223 are best-fit by the \textsc{MYTorus} model in the decoupled mode where the relative normalization between the transmitted and Compton scattered emission ($A_{\rm S}$) is fitted as a free parameter (see Appendix \ref{ruled_out_fits} for a discussion of \textsc{MYTorus} fit results that we rejected) and by the \textsc{borus02} model. The fitted parameters are summarized in Tables \ref{mytor_decoup_freeas_table} - \ref{borus_table}, and the fitted spectra and model components are shown in Figures \ref{mytor_decoup_freeas_plots} - \ref{borus02_plots}. The \textsc{borus02} model finds a slightly enhanced iron abundance ($A_{\rm Fe} = 2.2^{+1.8}_{-0.9}$) and the covering factor is not well constrained ($CF_{\rm tor} > 0.29$).

At first glance, the model fits imply contradictory results. The \textsc{MYTorus} scattered AGN model finds a much higher column density along the line of sight ($N_{\rm H,los} = 0.56^{+2.4}_{-1.9} \times 10^{24}$ cm$^{-2}$) than the global average column density ($N_{\rm H,global} = 0.07^{+0.03}_{-0.02} \times 10^{24}$ cm$^{-2}$), while the opposite is found from the \textsc{MYTorus} jet model ($N_{\rm H, los}$ = 0.08$^{+0.04}_{-0.02} \times 10^{24}$ cm$^{-2}$; $N_{\rm H, global}$ = 0.80$^{+0.48}_{-0.33} \times 10^{24}$ cm$^{-2}$) and \textsc{borus02} model ($N_{\rm H, los}$ = 0.16$^{+0.05}_{-0.03} \times 10^{24}$ cm$^{-2}$; $N_{\rm H, global}$ = 0.87$^{+1.13}_{-0.68} \times 10^{24}$ cm$^{-2}$). The models also disagree as to the dominant source of emission between $\sim 2-5$ keV, with the \textsc{MYTorus} scattered AGN model favoring a reflection-dominated spectrum while the \textsc{MYTorus} jet model finds a transmission dominated-spectrum. The \textsc{borus02} model is transmission-dominated above 3 keV but the scattered AGN component drives the spectrum below 3 keV.

The fitted power law slope of the AGN continuum from the \textsc{MYTorus} scattered AGN model, $\Gamma$ = 1.91$^{+0.33}_{-0.27}$, is typical of radiatively efficient, radio quiet AGN \citep[$\Gamma \sim 1.9$;][and references therein]{reeves2000}, while some past studies indicate that radio loud AGN tend to have flatter spectra ($\Gamma \sim 1.6$). The fitted AGN spectral slope for both the \textsc{MYTorus} jet model and \textsc{borus02} model are harder than that measured with the \textsc{MYTorus} scattered AGN model, but a clear dichotomy in the spectral hardness between radio quiet and radio loud AGN is not always observed \citep[e.g.,][]{sambruna2001,hardcastle2006,hardcastle2007,hardcastle2009}, meaning that we can not use spectral hardness alone to reject the \textsc{MYTorus} scattered AGN model. Furthermore, the best fit AGN spectral slope pegged at the minimum allowed value ($\Gamma_{\rm limit}$ = 1.4) in the \textsc{MYTorus} jet model and \textsc{borus02} model, raising questions about the reliability of these model fits.

Here is where the models agree: the obscuring medium reprocessing the X-ray emission from the AGN in 3C 223 is non-homogeneous, with up to an order of magnitude difference between the line-of-sight and global average column density. The column density of the obscuring gas is quite high, though not Compton-thick, reaching levels above 10$^{23}$ cm$^{-2}$. And a closer inspection of the fit parameters reveal that extra physics is at play to boost the soft X-ray spectrum beyond what can be accounted for in the radiative transfer codes used to define the \textsc{MYTorus} and \textsc{borus02} model grids. The normalization between the transmitted and Compton-scattered components of both \textsc{MYTorus} models is constrained to be several times greater than unity ($A_{\rm S} \sim 5-6$) and the AGN ``scattering fraction'' found by the \textsc{borus02} model ($f_{\rm scatt} = 27^{+5}_{-3}$\%) is much higher than the $<$1-3\% typically found in AGN \citep{winter2009,turner2009,lamassa2009,lamassa2011} and is unphysical \citep{buchner2019}. In fact, this high scattering fraction, and measured column densities, are consistent with those found from the \textsc{MYTorus} scattered AGN model using \textsc{MYTorus} in decoupled mode and with $A_{\rm S}$ frozen to unity (see Appendix \ref{ruled_out_fits}), so there is a degeneracy between accommodating the $<$3 keV part of X-ray spectrum with an unphysical scattering fraction or an elevated normalization between the transmitted and Compton-scattered emission. Since we allow iron abundance to be a free parameter in the spectral fitting, we can rule out enhanced iron abundances as a main driver for an elevated normalization between the Compton-scattered and transmitted emission.
 
To test whether excess unresolved thermal emission at soft energies could give rise to the high scattering fraction/elevated normalization, we included an \textsc{apec} component for this potential emission in our \textsc{MYTorus} scattered AGN model. The fit improved marginally ($\chi^2$=330.2 for 408 degrees of freedom), but the best fit $A_{\rm S}$ was even higher than when this component was not included ($A_{\rm S} = 14^{+15}_{-8}$). When adding the \textsc{apec} component to the \textsc{MYTorus} jet model, the \textsc{apec} model parameters were unconstrained. We thus conclude that high relative normalization between the transmitted and Compton scattered component is not due to unresolved host galaxy emission boosting the spectrum at soft energies.

\begin{deluxetable}{lll}[h]
\tablecaption{\label{mytor_decoup_freeas_table}Decoupled \textsc{MYTorus} Fit Parameters\tablenotemark{a}}
\tablehead{
  \colhead{Parameter} & \colhead{Scattered AGN} &  \colhead{Jet} 
}
\startdata
$\Gamma_{\rm AGN}$                         & 1.91$^{+0.33}_{-0.27}$ & $<$1.62 \\
Power law norm - AGN (10$^{-4}$)          & 6.05$^{+14.9}_{-3.95}$ & 0.57$^{+0.35}_{-0.10}$\\
$\Gamma_{\rm Jet}$                         & \nodata               & 1.83$^{+0.24}_{-0.23}$\\
Power law norm - Jet (10$^{-4}$)          & \nodata               & 0.26$\pm0.02$\\
$N_{\rm H,los}$ (10$^{24}$ cm$^{-2}$)       & 0.56$^{+0.24}_{-0.19}$  & 0.08$^{+0.04}_{-0.02}$ \\
$N_{\rm H,global}$ (10$^{24}$ cm$^{-2}$)    & 0.07$^{+0.03}_{-0.02}$  & 0.80$^{+0.48}_{-0.33}$\\
$A_{\rm S}$\tablenotemark{b}              & 6.4$^{+3.8}_{-3.1}$ & 5.1$^{+2.3}_{-1.8}$ \\
$f_{\rm scatt}$ (10$^{-2}$)                & 3.9$^{+7.4}_{-2.8}$ & \nodata \\
$\chi^2$ (dof)                           & 338.8 (395) & 336.7 (394) \\
\hline
\multicolumn{3}{c}{X-ray luminosities\tablenotemark{c}} \\
\hline
Log($L_{\rm 2-10keV,observed}$ erg s$^{-1}$)  & 43.27$\pm0.14$ & 43.26$^{+0.21}_{-0.08}$ \\
Log($L_{\rm 10-40keV,observed}$ erg s$^{-1}$) & 43.70$\pm0.14$ & 43.79$^{+0.21}_{-0.08}$ \\
Log($L_{\rm 2-10keV,intrinsic}$ erg s$^{-1}$) & 43.87$^{+0.74}_{-0.69}$ & 43.22$^{+0.21}_{-0.25}$\\
\enddata
\tablenotetext{a}{Best fit parameters from modeling the {\it XMM-Newton} and {\it NuSTAR} spectra of 3C 223 with the \textsc{MYTorus} model in decoupled mode, where the line-of-sight column density ($N_{\rm H,los}$) of the X-ray obscurer is fitted independently from the global average column column density ($N_{\rm H,global}$). The soft emission (0.5 - 2 keV) was modeled assuming it originates from AGN emission scattered into our line of sight (``Scattered AGN'' column), or that it emanates from the base of an X-ray jet (``Jet'' column), with a powerlaw spectral index ($\Gamma_{\rm Jet}$) independent of the primary AGN component ($\Gamma_{\rm AGN}$).}
\tablenotetext{b}{Normalization between the transmitted AGN continum and the Compton scattered and fluorescent line emission.}
\tablenotetext{c}{X-ray luminosities refer to the AGN luminosities and are derived from the average X-ray fluxes measured by the spectral model for each instrument. The errors on the observed luminosities are based on propogating errors in the powerlaw normalization. The intrinsic luminosity is calculated from an absorption-corrected powerlaw model with $\Gamma$ and the normalization set to the best-fit values from spectral fitting. The errors on the intrinsic luminosity account for the range of possible luminosities based on the errors in the normalization and spectral index of the powerlaw model.}
\end{deluxetable}

\begin{deluxetable}{lll}[h]
\tablecaption{\label{borus_table}\textsc{borus02} Fit Parameters}
\tablehead{
  \colhead{Parameter} & \colhead{Value} \\
}
\startdata
$\Gamma$                            &  $<$1.50 \\
Power law norm (10$^{-4}$)          &  0.69$^{+0.02}_{-0.11}$ \\
Inclination angle ($^{\circ}$)       & $>$55 \\
$N_{\rm H,los}$ (10$^{24}$ cm$^{-2}$)    & 0.16$^{+0.05}_{-0.03}$ \\
$N_{\rm H,global}$ (10$^{24}$ cm$^{-2}$) & 0.87$^{+1.13}_{-0.68}$ \\
$CF_{\rm tor}$\tablenotemark{a} & $>$0.29 \\
$A_{\rm Fe}$\tablenotemark{b} & 2.2$^{+1.8}_{-0.9}$\\
$f_{\rm scatt}$ (10$^{-2}$)   & 27$^{+5}_{-3}$ \\
$\chi^2$ (dof)            & 302.1 (348) \\
\hline
\multicolumn{2}{c}{X-ray luminosities} \\
\hline
Log($L_{\rm 2-10keV,observed}$ erg s$^{-1}$)  & 43.26$^{+0.06}_{-0.01}$ \\
Log($L_{\rm 10-40keV,observed}$ erg s$^{-1}$) & 43.79$^{+0.06}_{-0.01}$ \\
Log($L_{\rm 2-10keV,intrinsic}$ erg s$^{-1}$) & 43.30$^{+0.01}_{-0.15}$ \\
\enddata
\tablenotetext{a}{$CF_{\rm tor}$ refers to the torus covering factor which is defined as cos($\theta_{\rm tor}$), where $\theta_{\rm tor}$ is the torus opening angle. $C_{\rm tor}$ can range from 0.1 (low covering factor) to 1.0 (spherical coverage).}
\tablenotetext{b}{$A_{\rm Fe}$ is the fitted iron abundance relative to solar.}
\end{deluxetable}

\section{Discussion}

\subsection{Insight into the Obscuring Medium: Inhomogeneity and Physical Implications of Deviations from Model Assumptions}\label{obsc_med}
3C 223 joins the list of obscured AGN whose obscuring medium is shown via broad band X-ray spectral coverage (0.5 - 30 keV), where enough counts are garnered to permit use of physically motivated spectral models, to be more complicated than that assumed by simplistic models where the torus is homogeneous and uniform \citep{yaqoob2015,balokovic,tzanavaris2021}. The X-ray spectra of 3C 223 require the presence of heavy attenuation ($>10^{23}$ cm$^{-2}$) along with gas clouds with column densities up to an order of magnitude lower, though how this gas is distributed depends on the spectral model. Studies that purport to trace the evolution of obscured supermassive black hole accretion by analyzing the unresolved cosmic X-ray background rely on models that assume a uniform column density for all sight-lines \citep{gilli,treister,akylas,ueda,ananna}, oversimplifying our emergent picture of the inhomogeneous reprocessing medium in AGN where the global column density may be the more relevant parameter \citep[see][]{yaqoob2015}.

\begin{figure*}[h]
  \begin{center}
    \includegraphics[scale=0.25]{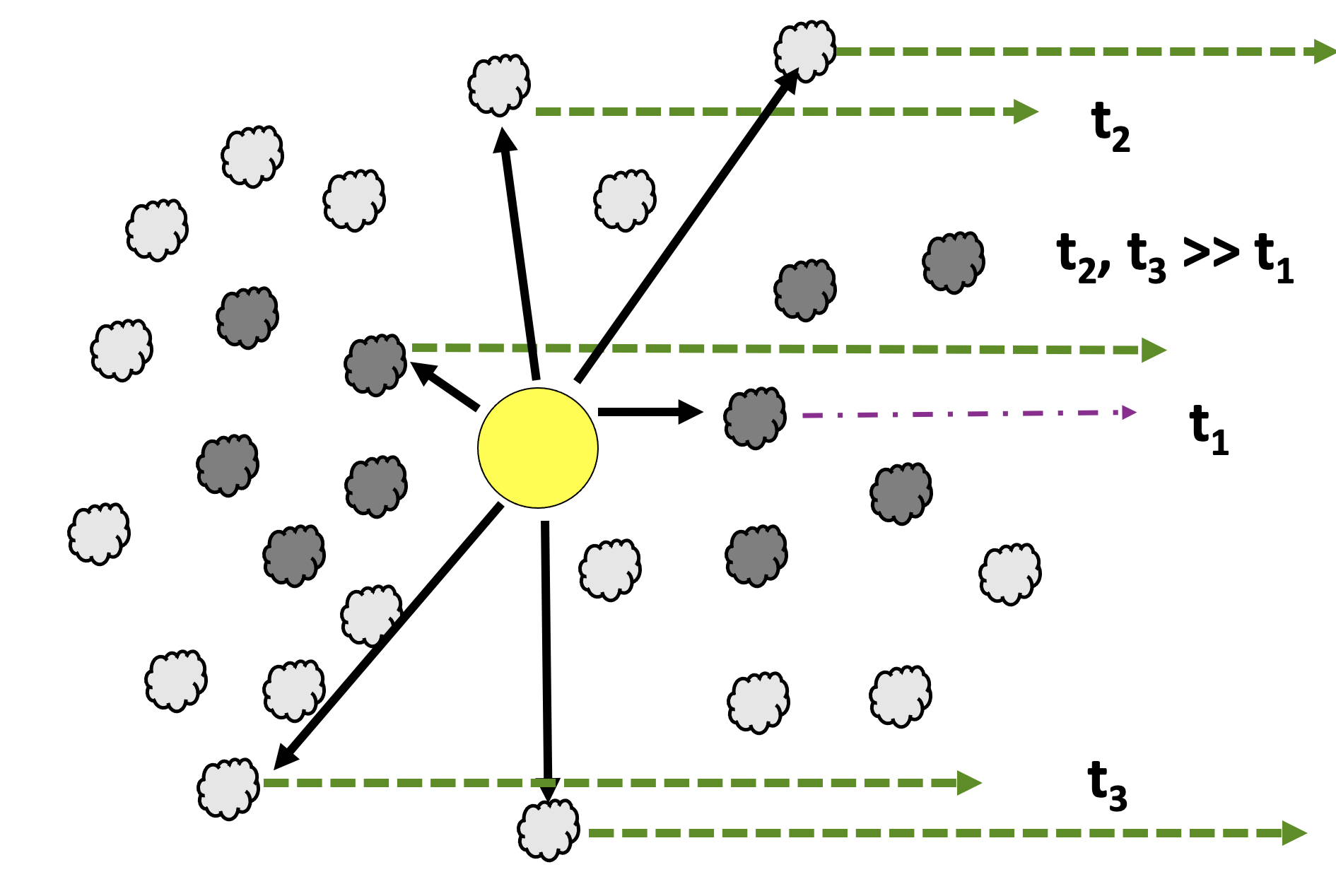}~
    \hspace{0.3cm}
    \includegraphics[scale=0.25]{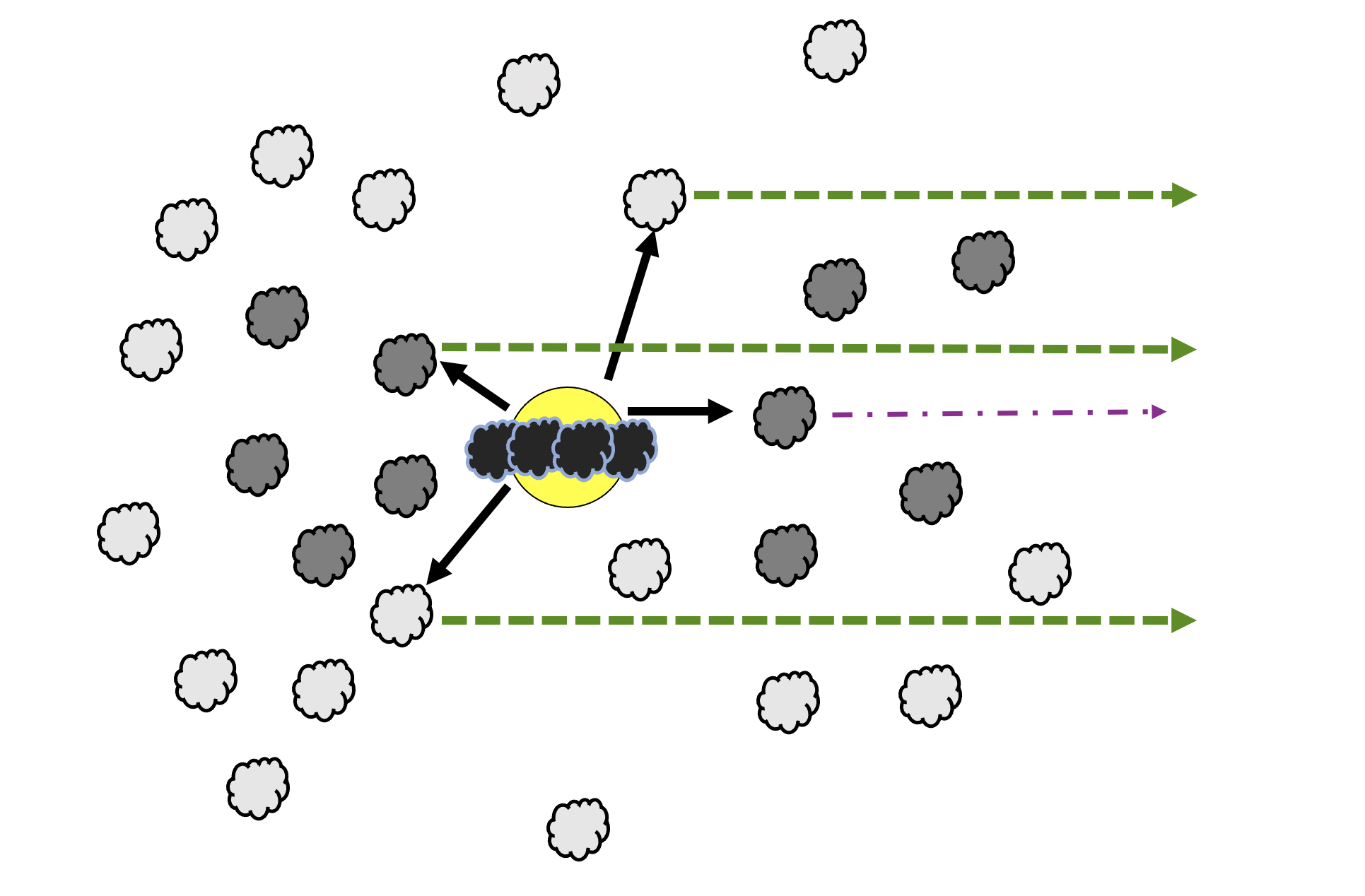}\\
  \includegraphics[scale=0.25]{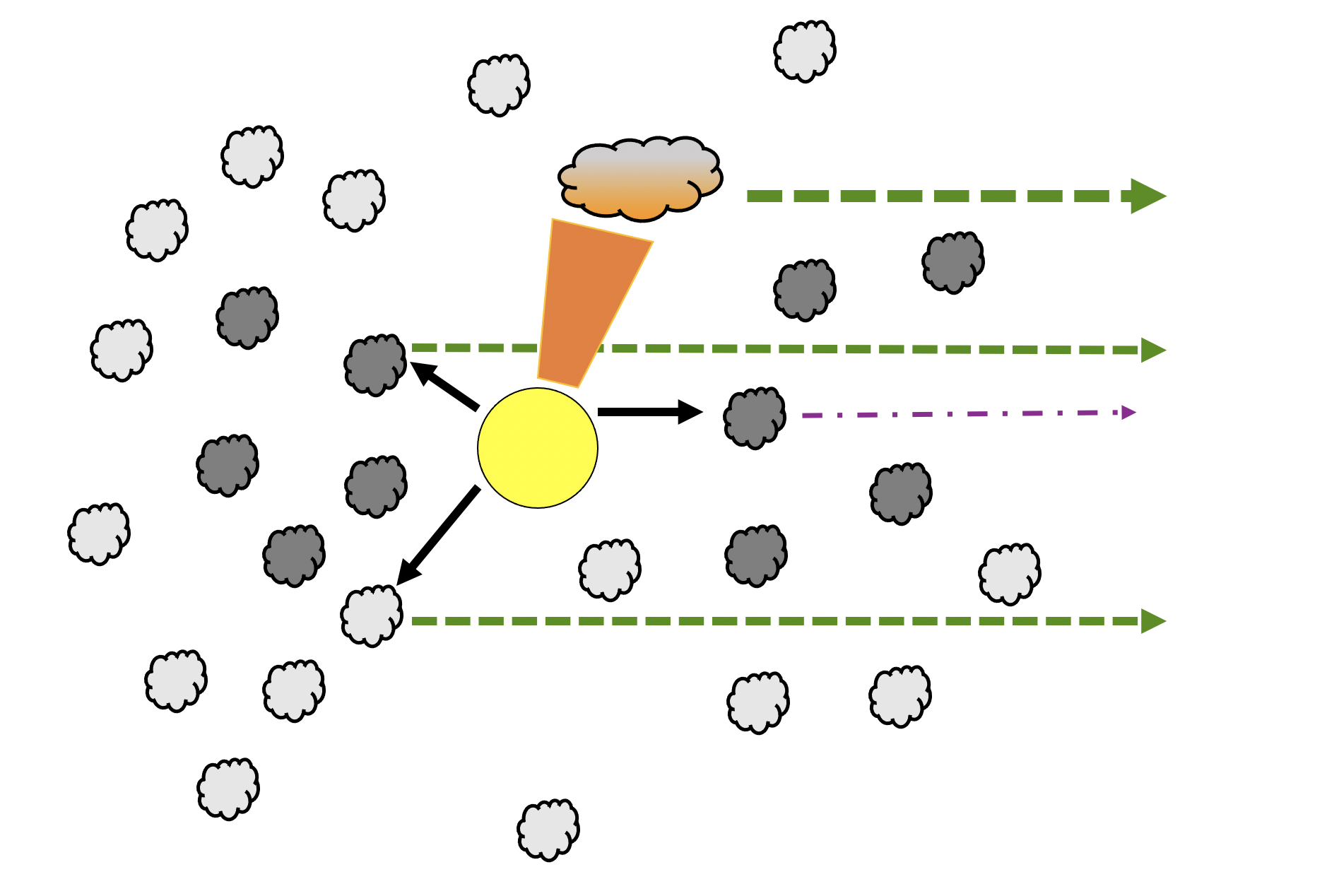}
  \caption{\label{geometry-sketches} Illustrations of possible geometries that could give rise to an elevated $A_{\rm s}$ or $f_{\rm scatt}$ value. In all cases, the yellow circle marks the X-ray emitting source, the dark grey clouds represent high column density gas clouds ($>10^{23}$  cm$^{-2}$),  the light grey clouds indicate lower column density gas clouds (several $\times 10^{22}$ cm$^{-2}$), and the observer is on the right side of the panel. The black solid arrows represent direct emission from the X-ray source, the dash-dotted purple line denotes the zeroth-order continuum after being absorbed by the line-of-sight gas, and the green dashed lines illustrate the X-rays that have been Compton-scattered by the global distribution of clouds, producing the Fe K$\alpha$ line emission and continuum features of X-ray reprocessing. {\it Top left}: Here the Compton-scattering region lies many parsecs away from the X-ray source while the line-of-sight clouds are much closer in, such that the time delay between observing the Compton-scattered photons ($t_{2}$, $t_{3}$) and absorbed line-of-sight photons ($t_{1}$) is much greater than the time period of the observations. {\it Top right}: In this scenario, the X-ray emitting region is obscured by a patchy equatorial ring of heavily Compton-thick material ($N_{\rm H} > 10^{26}$ cm$^{-2}$, black clouds in diagram), allowing only a fraction of light to intercept the lower column density line-of-sight clouds while the global distribution of matter has a largely unimpeded view of the X-ray source. {\it Bottom}: A jet beamed in a direction not along the line of sight (shown in orange), can impart significant energy to the cloud that it illuminates (orange colored cloud), boosting the production of Fe K$\alpha$ photons (thicker green dashed line) compared with those generated in the line-of-sight obscuring clouds.}
  \end{center}
\end{figure*}

In the \textsc{MYTorus} model, an $A_{\rm S}$ value near unity implies that the intrinsic AGN continuum is constant, or that the reprocessor responds to any changes in the continuum on time scales that are much less than the spectrum integration time. A value that deviates significantly from unity, and the high scattering fraction in the \textsc{borus02} model, indicate that additional physical processes are impacting the observed spectrum, with some possibilities illustrated in Figure \ref{geometry-sketches}.

Variability can be one such cause. The 0.5-10 keV X-ray observations were taken between the years of 2001 through 2019, with no variability observed in these spectra (Appendix \ref{xray_var}). If time delays are responsible for the high $A_{\rm S}$ value, that could indicate that the intrinsic continuum might have been much brighter years before the first {\it XMM-Newton} observing epoch and that the reprocessing material has not yet responded to this change. Such a scenario, illustrated in Figure \ref{geometry-sketches} (top left), would indicate that the physical scale of the Compton-scattering region is larger than the light-travel distance covered by the 18 year window of the {\it XMM-Newton} observations ($\sim$5.5 pc, which is larger than the putative torus). Detailed studies of nearby obscured AGN show evidence that though the bulk of the emission related to X-ray reprocessing (e.g., Fe K$\alpha$ emission) is nuclear \citep[e.g.,][]{gandhi2015}, a fraction of this emission is detected on scales beyond the several pc torus \citep[e.g., NGC1068, ESO 428-G014, NGC 5643, NGC 7212, NGC 4388;][]{bauer2015,fabbiano2017,fabbiano2018,jones2020}. With our chosen cosmology, the physical scale at $z=0.1365$ is 2.5 kpc/$^{\prime\prime}$, so our X-ray spectra sample physical sizes of $\sim75-110$ kpc. The high $A_{\rm S}$ or $f_{\rm scatt}$ value could then indicate that a fraction of the Compton scattering is occuring within a region well within the $\sim$100 kpc sampled by our X-ray spectra, but on scales larger than the light-travel distance of 5.5 pc (otherwise we would have observed spectral variability in the intrinsic continuum between {\it XMM-Newton} epochs or measured a relative normalization closer to unity).

Alternatively, the covering factor of the obscuring medium in 3C 223 may deviate significantly from 0.5. The X-ray spectral shape supports a geometry where the line of sight is obscured by just enough material to boost the Fe K$\alpha$ EW \citep[EW $\sim 0.5 \pm 0.2$ keV,][]{jia} while not imparting a spectral curvature between 2 - 6 keV associated with Compton-thick obscuration. As suggested in \citet{lamassa2014} and illustrated in the top right panel of Figure \ref{geometry-sketches}, perhaps there is a ring of clouds around the X-ray emitting source that are heavily Compton-thick (i.e., $N_{\rm H} > 10^{26}$ cm$^{-2}$) blocking most of the light in the equatorial direction while the global distribution of clouds can ``see'' most of the X-ray source. A fitted normalization of $\sim$5-6 between the Compton-scattered/line emission and transmitted continuum can indicate that the matter along the line of sight is illuminated by $\sim$20\% of the X-ray luminosity that is intercepted and reprocessed by the global medium. We attempted to model the X-ray spectra with \textsc{UxClumpy} to 1) test for the presence of an inner Compton-thick ($N_{\rm H} > 10^{25}$ cm$^{-2}$) ring of material and 2) measure the angular width of the obscuring clouds. However, the inner Compton-thick ring is constrained to be smooth so this model is unable to test our posited scenario as this requires a patchy ring with a covering factor of $\sim$80\%. Our modeling with \textsc{UxClumpy} resulted in a fit that did a poorer job of accommodating the Fe K$\alpha$ complex compared with our best-fit \textsc{MYTorus} model (Appendix \ref{uxclumpy_fitting}). And similar to the \textsc{borus02} model, \textsc{UxClumpy} modeling results in an unphysical scattering fraction with a best-fit $f_{\rm scatt}$ of 100\%. 

A final possibility may be ascribed to interactions between the jet and the putative torus. If the jet is beamed in a direction not along the line of sight, then the global matter distribution intercepts a larger fraction of the intrinsic AGN energy (Figure \ref{geometry-sketches}, bottom), boosting the production of Fe K$\alpha$ line photons compared with those created along the line of sight \citep[see][]{yaqoob1993,yaqoob1999}. This intrinsic anisotropy will cause the transmitted continuum to appear to be suppressed relative to the global medium producing the bulk of the Fe K$\alpha$ emission, manifesting in the large Fe K$\alpha$ EW visible in the X-ray spectrum. Thus the relative normalization between the Compton scattered/fluorescent line component and transmitted component will be constrained to values greater than unity. Examples of radio jet interactions with the AGN obscuring medium have been previously observed: in-depth investigations of compact symmetric objects (radio loud AGN whose radio emission is contained within 1 kpc of the nucleus) reveal that the most compact of these radio objects reside in AGN with high column densities ($N_{\rm H} > 10^{23}$ cm$^{-2}$), which suggests that higher density gas can restrict the expansion or production of larger scale radio jets \citep{sobolewska2019}.

\subsection{Consistency between X-ray and Mid-Infrared Emission}
The dusty torus beyond the accretion disk and corona absorbs optical-ultraviolet photons and re-radiates this emission at mid-infrared (MIR) wavelengths. Coupling between the accretion disk and X-ray emitting corona would then impose a correlation between the mid-infrared and X-ray emission, which has been observed in a number of studies \citep{lutz,ramosalmeida,horst2008,gandhi2009,levenson2009,hoenig2010,stern2015}. Taken a step further, observed X-ray emission that is heavily suppressed relative to the MIR emission is sometimes used as a proxy of Compton thick obscuration \citep{daddi2007,lamassa2009,lamassa2011,bauer2010,donley}, though the X-ray to mid-infrared ratio can be influenced more by the spectral shape of the X-ray continuum and covering factor of the torus than the column density \citep{yaqoob2011}.

Using the observed 12$\mu$m flux of 3C 223 from the $W3$ magnitude of the All-{\it WISE} survey \citep{wright,cutri}, where $W3$ = 7.8 (Vega), we can use the MIR-X-ray relation from \citet{asmus} to estimate the intrinsic 2-10 keV luminosity. We find a predicted X-ray luminosity of 1.61 $\times 10^{44}$ erg s$^{-1}$, which is about twice the value we derived from fitting the X-ray spectra with the \textsc{MYTorus} scattered AGN model ($L_{\rm 2-10keV,instrinsic} = 7.4 \times 10^{43}$ erg s$^{-1}$). We note that this $L_{\rm 12\mu m} - L_{\rm 2-10 keV}$ relation is derived using high-resoluton MIR measurements that isolate the galaxy cores, minimizing contamination from host galaxy star formation. The lower resolution {\it WISE} measurement encompasses the whole galaxy, which could be partially responsible for the higher predicted X-ray luminosity compared with our X-ray spectral measurements. Still, the agreement between the two values is within the scatter of the \citet{asmus} relation ($\sigma \sim 0.3$ dex), which is consistent with a physical link between the powering of the corona and heating of the torus.

\subsection{Bolometric Output and Accretion Rate of the 3C 223 Black Hole}\label{edd}
The Eddington luminosity ($L_{\rm Edd} = 1.26 \times 10^{38}$ ($M_{\rm BH}$/$M_{\rm \odot}$) erg s$^{-1}$) represents the limit at which matter can accrete onto the central supermassive black hole. The Eddington ratio, defined as the ratio of the bolometric AGN luminosity ($L_{\rm bol}$) to the Eddington luminosity ($\lambda_{\rm Edd} = L_{\rm bol}/L_{\rm Edd}$), parameterizes the black hole accretion rate.

Since the line of sight to the accretion disk is obscured in Type 2 AGN, we are unable to calculate the black hole mass using virial mass estimators that relate the width of broad emission lines to the accretion disk luminosity. We thus use the empirical correlation between the black hole mass and velocity dispersion of stars in the host galaxy \citep[i.e., $M_{\rm BH}$-$\sigma$;][]{ferrarese} to estimate $M_{\rm BH}$ for the AGN in 3C 223. From a measured velocity dispersion of $\sigma$ = 202 km s$^{-1}$ \citep{bettoni}, \citet{woo} used the \citet{tremaine} $M_{\rm BH}$-$\sigma$ relation to calculate a black hole mass of Log($M_{\rm BH}/M_{\rm \odot}$) = 8.15 dex for 3C 223. $L_{\rm Edd}$ is thus $1.78 \times 10^{46}$ erg s$^{-1}$.

The AGN bolometric luminosity can be measured by integrating the AGN-only contribution to the multi-wavelength spectral energy distribution (SED). This procedure requires panchromatic coverage from rest-frame ultraviolet through mid-to-far infrared emission, and faces challenges from accurately decomposing host-galaxy from AGN emission (especially in Type 2 AGN). Alternate methods of estimating the bolometric luminosity involve measuring a quantity that traces the intrinsic AGN emission and applying a correction to extrapolate to the total AGN luminosity. We have two such intrinsic AGN indicators at our disposal: the optical [\ion{O}{3}] line and the absorption-corrected 2-10 keV luminosity ($L_{\rm 2-10keV,intrinsic}$). We use both indicators and two different bolometric corrections to $L_{\rm 2-10keV,intrinsic}$ to estimate a range of possible Eddington ratios for 3C 223 that we report in Table \ref{agn_params}.

[\ion{O}{3}] forms in the AGN narrow line region (NLR), hundreds of parsecs beyond the obscuring torus, and is primarily ionized by accretion disk photons, making it a reasonable proxy of the intrinsic AGN emission \cite[e.g.,][]{mulchaey,kauffmann,lamassa2010}. \citet{liu2009} calibrated the bolometric correction to $L_{\rm [O\,III]}$ for Type 2 quasars from SDSS using:
\begin{equation}
  {\rm Log}\left(\frac{L_{\rm bol}}{L_{\rm \odot}}\right) = 0.99 \times \rm{Log}\left(\frac{L_{\rm [O\,III]}}{L_{\rm \odot}}\right) + 3.5.
\end{equation}
They quote an uncertainty of 0.5 dex in $L_{\rm bol}$ for this relation. We note that one factor that can contribute to this uncertainty is reddening towards the narrow line region which can attenuate the observed [\ion{O}{3}] emission \cite[see, e.g.,][]{lamastra}. On the other hand, local ionization from jet heating can power the optical emitting line region \citep{holt, nesvadba}, meaning that the [\ion{O}{3}] line may not cleanly trace AGN ionization in radio loud sources \citep[see, e.g., 4C+29.30;][]{vanbreugel, siemiginowska, couto}. With these caveats in mind, the observed Log($L_{\rm [O\,III]}/L_{\rm \odot}$) value of 8.78 dex \citep{reyes} indicates an [\ion{O}{3}]-determined bolometric AGN luminosity of $L_{\rm bol,[O\,III] } = 5.93 \times 10^{45}$ erg s$^{-1}$, for an Eddington ratio of $\lambda_{\rm Edd,[O\,III]}$ = 0.33.

The form of the X-ray bolometric correction likely depends on bolometric AGN luminosity \citep{marconi,hopkins} or Eddington ratio \citep{vasudevan2007,lusso2010}. Using the relations for the bolometric correction calibrated on Type 2 AGN from the {\it XMM}-COSMOS survey from \citet{lusso2012}, we estimate $L_{\rm bol}$ and $\lambda_{\rm Edd}$ from our measured intrinsic 2-10 keV X-ray luminosity derived via the \textsc{MYTorus} scattered AGN model ($L_{\rm 2-10keV,intrinsic} = 7.4\times 10^{43}$ erg s$^{-1}$, Table \ref{mytor_decoup_freeas_table}). We calculate these parameters using both the calibrated $L_{\rm bol}$-dependent and $\lambda_{\rm Edd}$-dependent bolometric correction relations presented in \citet{lusso2012}.

The functional form of the $L_{\rm bol}$-dependent bolometric correction is:
\begin{equation}
  {\rm Log}\left(\frac{L_{\rm bol}}{L_{\rm 2-10keV}}\right) = 0.23x\, +\, 0.05x^2\, +\, 0.001x^3\, +\, 1.256,
\end{equation}
where $x$ = Log($L_{\rm bol}/L_{\rm \odot}$) - 12. We solved for $L_{\rm bol}$ numerically given our intrinsic X-ray luminosity measurement, finding $L_{\rm bol,2-10 keV}$ = $1.0 \times 10^{45}$ erg s$^{-1}$ with $\lambda_{\rm Edd,2-10keV}$ = 0.06.

The calibrated $\lambda_{\rm Edd}$-dependant bolometric correction for Type 2 AGN is given by:
\begin{equation}
  {\rm Log}\left(\frac{L_{\rm bol}}{L_{\rm 2-10keV}}\right) = 0.621x\, +\, 1.947,
\end{equation}
where $x$ = Log($\lambda_{\rm Edd}$) \citep{lusso2012}. This relation gives $\lambda_{\rm Edd}$ = 0.07 and a bolometric luminosity of 1.3$\times10^{45}$ erg s$^{-1}$.

Taken together, and given the uncertainties in the relationships and the measured values, we conclude that the AGN in 3C 223 is accreting at greater than 5\% Eddington. This accretion rate is consistent with that of other high excitation radio galaxies (HERG), pointing to radiatively efficient accretion described by the standard \citet{shakura} optically thick, geometrically thin accretion disk model \citep{best2012}. We note that \citet{hardcastle2007} proposed that major galaxy mergers can provide a resevoir of cold gas for powering radiatively efficient accretion in HERGs, but {\it Hubble Space Telescope} images of 3C 223 do not reveal morphological evidence of recent merger activity \citep{madrid}.

\begin{deluxetable}{lll}[h]
\tablecaption{\label{agn_params}3C 223 AGN Physical Parameters\tablenotemark{a}}
\tablehead{
  \colhead{Parameter} & \colhead{Value} &  \colhead{Reference} \\
}
\startdata
Log($M_{\rm BH}/M{\rm \odot}$)           & 8.15 dex & \citet{woo} \\
$L_{\rm Edd}$ (erg s$^{-1}$)              & $1.78 \times 10^{46}$  & \nodata \\
$L_{\rm [O\,III]}$(erg s$^{-1}$)                         & $2.29 \times 10^{42}$  & \citet{reyes} \\
$L_{\rm [O\,III],intrinsic}$(erg s$^{-1}$)\tablenotemark{b} &  $5.81\times10^{42}$    & \nodata \\
$L_{\rm bol,[O\,III] }$ (erg s$^{-1}$)\tablenotemark{c}    & $5.93 \times 10^{45}$ & \citet{liu2009} \\
$\lambda_{\rm Edd,[O\,III]}$              & 0.33 & \nodata  \\
$L_{\rm bol,2-10keV}$\tablenotemark{d}     & $1.0 \times 10^{45}$ & \citet{lusso2012} \\
$\lambda_{\rm Edd,2-10keV}$\tablenotemark{d} & 0.06 & \nodata \\
$L_{\rm bol,2-10keV}$\tablenotemark{e}       & $1.3 \times 10^{45}$ & \citet{lusso2012} \\
$\lambda_{\rm Edd,2-10keV}$\tablenotemark{e} &  0.07 & \nodata \\
$L_{\rm jet}$ (erg s$^{-1}$)\tablenotemark{f}  & 2.1 $\times 10^{45}$ & \nodata \\
\enddata
\tablenotetext{a}{We estimated the AGN bolometric luminosity ($L_{\rm bol}$) and Eddington ratio ($\lambda_{\rm Edd}$) using the [\ion{O}{3}] emission line and intrinsic 2-10 keV luminosity. The reported subscripts on these parameters refer to the quantity used to derive $L_{\rm bol}$ and $\lambda_{\rm Edd}$. For $L_{\rm bol}$ and $\lambda_{\rm Edd}$ derived from $L_{\rm 2-10keV,intrinsic}$, we used two different functional forms of the bolometric correction. }
\tablenotetext{b}{Extinction-corrected [\ion{O}{3}] luminosity. See Section \ref{qso_comp} and Appendix \ref{o3corr_derivation} for details.}
\tablenotetext{c}{$L_{\rm [O\,III]}$ measured from \citet{reyes} and bolometric correction to $L_{\rm [O\,III]}$ from \citet{liu2009}. Note that jet heating could contribute to the [\ion{O}{3}] line luminosity in radio loud AGN \citep[e.g.,][]{vanbreugel, couto}.}
\tablenotetext{d}{Using the $L_{\rm bol}$-dependent bolometric correction relation for Type 2 AGN from \citet{lusso2012} and $L_{\rm 2-10keV,intr} = 7.4\times10^{43}$ erg s$^{-1}$ from the \textsc{MYTorus} scattered AGN model.}
\tablenotetext{e}{Using the $\lambda_{\rm Edd}$-dependent bolometric correction relation for Type 2 AGN from \citet{lusso2012} and $L_{\rm 2-10keV,intr} = 7.4\times10^{43}$ erg s$^{-1}$ from the \textsc{MYTorus} scattered AGN model.}
\tablenotetext{f}{Jet luminosity using $P_{\rm mech, cavity} =  3 \times 10^{37} (L_{\rm 1.4 GHz}/10^{25}{\rm W Hz^{-1}})^{0.68}$ W \citep{heckmanbest}, and $L_{\rm 1.4 GHz}$ =  3.54 Jy \citep{kuzmicz}}
\end{deluxetable}

\subsection{Radio Properties}\label{radio}
Typical of FR II radio galaxies, VLA images of 3C 223 at 20 cm and 8.4 GHz depict a radio core at the center of the AGN with extended radio lobes that span about 1.5-2 arcminutes to the Northwest and Southeast directions \citep{baum,massaro2012}. The radio lobes extend in declination from approximately 35:51:30 $< \delta <$ 35:53:00 and 35:54:45 $< \delta <$ 35:56:30 (J2000 coordinates). We extracted the X-ray spectra from a circular aperture centered at the X-ray centroid of the AGN core ($\alpha$ = 9:29:52.95, $\delta$ = 35:53:58.5) with a 30$^{\prime\prime}$ radius aperture for {\it XMM-Newton} and a 45$^{\prime\prime}$ radius aperture for {\it NuSTAR}. The X-ray emission we analyzed is spatially coincident with the radio core but does not have contributions from the extended lobes. As noted previously, we took care to sample the background from a region that did not overlap with the radio lobes.

We compare the amount of energy released by the radio jet in the core of the galaxy with the AGN radiative power estimated in the previous section. The jet kinetic power can be quantified using $P_{\rm mech, cavity} =  3 \times 10^{37} (L_{\rm 1.4 GHz}/10^{25}{\rm W Hz^{-1}})^{0.68}$ W \citep{heckmanbest}. 3C 223 has a measured 1.4 GHz flux density of 3.54 Jy \citep{kuzmicz}, which gives a jet luminosity of $2.1 \times 10^{45}$ erg s$^{-1}$. This value is 12\% of our calculated Eddington luminosity, exceeding the intrinsic X-ray luminosity we measured, and is comparable to our estimates of the radiative bolometric luminosity. 

\subsection{3C 223 in Type 2 Quasar Parameter Space}\label{qso_comp}

The incidence of Compton-thick obscured quasars, the most luminous AGN, is of particular interest since some studies show a clear decline in AGN obscuration as luminosity increases \citep[the ``receding torus`` model,][]{lawrence1982,ueda2003,merloni2014,brightman2014}, though this could be due to the difficulty in finding the most heavily obscured AGN \citep[e.g.,][]{mateos2017}, and if there is a physical connection, the accretion rate rather than the AGN luminosity may be the driver \citep{winter2009,ricci2017}. Certainly a better census of the obscuration levels in the most luminous AGN would shed light on whether the apparent receding torus phenomenon is an observational bias or a physical effect, and if the latter, the roles that accretion rate and AGN energy output may play in shaping the circumnuclear environment.

Significant efforts have been made to reliably identify Type 2 quasars and then assess their obscuring column densities to identify the Compton-thick population. One method for identifying Compton-thick AGN relies on selecting obscured AGN on the basis of intrinsic AGN luminosity proxies that are relatively unaffected by the amount of circumnuclear obscuration and then examining the X-ray properties of these sources to search for evidence of Compton-thick obscuration \citep{bassani1999,heckman2005,panessa2006,lamassa2009,lamassa2011,lamassa2014,jia}. As noted above, the [\ion{O}{3}] line is one commonly used proxy of intrinsic AGN emission used to select samples of obscured AGN.

\citet{reyes} selected a sample of Type 2 quasars from SDSS that have optical emission line ratios consistent with AGN photionization and [\ion{O}{3}] luminosities exceeding $L_{\rm[O\,III]} > 10^{8.3} L_{\rm \odot}$ \citep[$7.64 \times 10^{41}$ erg s$^{-1}$;][]{zakamska}. 3C 223 (designated as SDSS J093952.74+355358.0) was included in this sample and was subsequently analyzed by \citet{jia} as part of an effort to analyze the X-ray spectra of [\ion{O}{3}]-selected Type 2 quasars to estimate the Compton-thick fraction.

Nine other [\ion{O}{3}]-selected Type 2 quasar candidates have been observed by {\it NuSTAR} from the \citet{reyes} and \citet{jia} sample \citep[see also][]{vignali2006,vignali2010}: Mrk 34 \citep[$z = 0.051$;][]{gandhi} and eight additional sources reported in \citet{lansbury2014} and \citet{lansbury2015} that lie in the redshift range between $0.094 < z < 0.49$. Of these nine, only Mrk 34 and one quasar from \citet[]{lansbury2015} (SDSS J121839.40+470627.7, $z=0.094$) were detected with adequate numbers of counts with {\it NuSTAR} to permit spectral fitting with the \textsc{MYTorus} model. Both are confirmed to be Compton-thick based on measuring their column densities and have intrinsic 2-10 keV X-ray luminosities computed. \citet{lansbury2014,lansbury2015} infer heavy obscuration in the remaining [\ion{O}{3}]-selected Type 2 quasars by comparing $L_{\rm 2-10keV}/L_{\rm 6 \mu m}$ and $L_{\rm 10-40keV}/L_{\rm 6 \mu m}$ to the X-ray - mid-infrared relation found for local AGN \citep{lutz}.\footnote{$L_{\rm 6 \mu m}$ refers to the rest-frame, reddening-corrected 6 $\mu$m luminosity.}

To the list of SDSS [\ion{O}{3}]-selected Type 2 quasars, we include Type 2 AGN from the 70-month Swift-BAT hard X-ray ($>$ 10 keV) survey \citep{baumgartner} whose [\ion{O}{3}] luminosities exceed the $L_{\rm[O\,III]} > 10^{8.3} L_{\rm \odot}$ quasar threshold defined in \citet{zakamska} and applied in \citet{reyes}. Like all X-ray surveys, the Swift-BAT survey has lower sensitivity to AGN that are Compton-thick along the line of sight \citep[e.g.,][]{lamassa2010}, but our goal here is to augment the comparison sample of Type 2 quasars observed with {\it NuSTAR}. As {\it NuSTAR} has been conducting a legacy survey of {\it Swift} BAT sources since 2012 \citep{alexander2013}, this dataset provides us a larger baseline for comparison.

Using emission line flux measurements from the BAT AGN Spectroscopic Survey \citep[BASS;][]{koss}, we identified 21 Type 2 AGN that can be classified as quasars on the basis of their [\ion{O}{3}] luminosity. Of these, three were observed by {\it NuSTAR} and their broad-band X-ray spectrum (0.5 - 80 keV) fitted with \textsc{MYTorus} by \citet{marchesi2018}: MCG +08-03-018 ($z=0.021$), CGCG 420-015 ($z=0.030$), and NGC 3393 ($z=0.013$). \citet{marchesi2019} measure a Compton-thick line-of-sight column density for NGC 3393, and heavily obscured, but Compton-thin column densities for the other two AGN \citep[$\sim5-8 \times10^{23}$ cm$^{-2}$, but see][for measured Compton-thick $N_{\rm H}$ values when using the \textsc{borus2} X-ray spectral model \citep{balokovic} with a free covering factor]{marchesi2019}. From these fits, \citet{marchesi2018} calculated absorption-corrected 2-10 keV luminosities that we use below.

\begin{figure}[h]
  \begin{center}
    \includegraphics[scale=0.5]{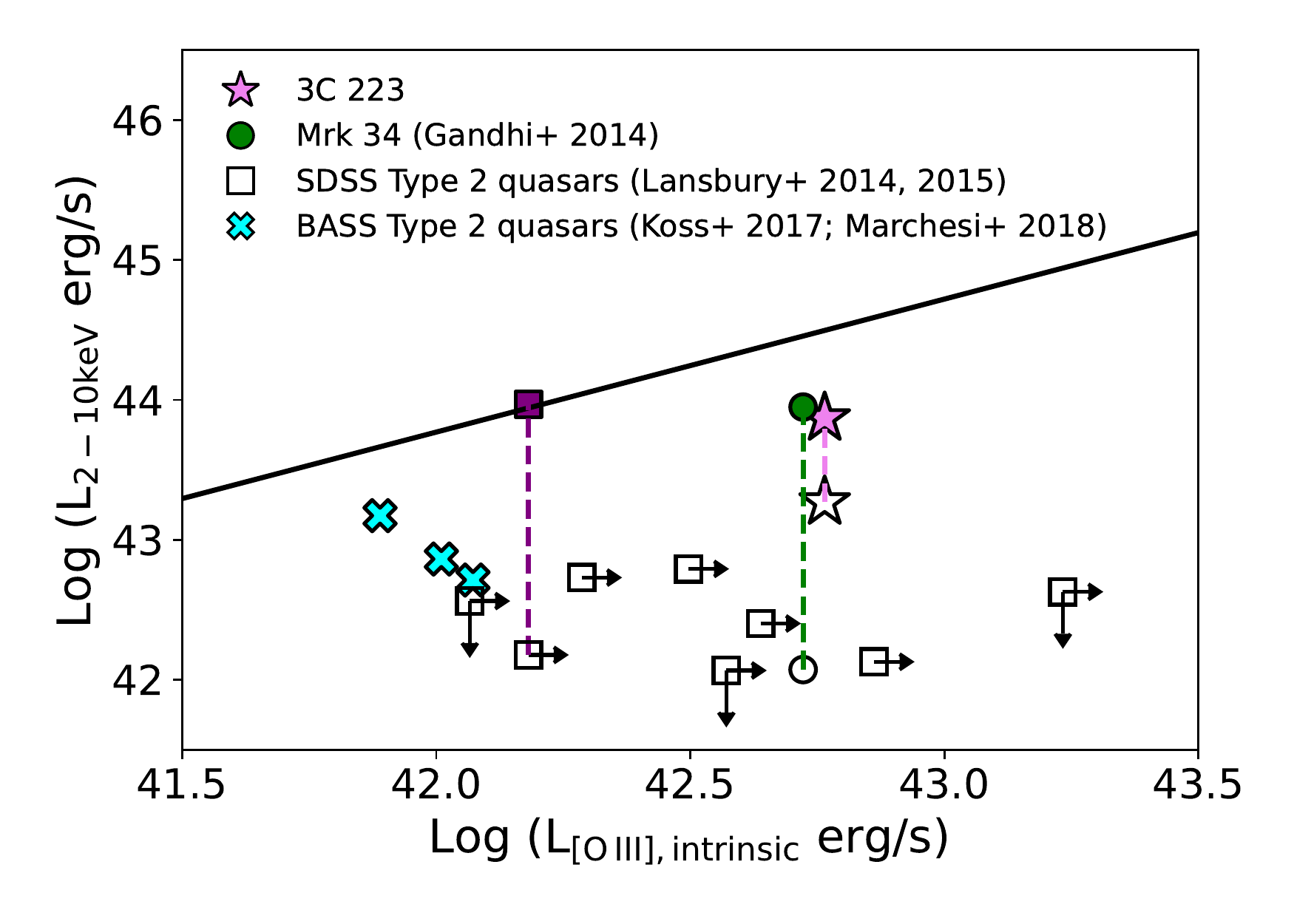}
    \caption{\label{lo3_v_lx} X-ray luminosity (2-10 keV) as a function of [\ion{O}{3}] luminosity for 3C 223 and other [\ion{O}{3}]-selected Type 2 quasars observed by {\it NuSTAR}. Filled data points represent the absorption-corrected 2-10 keV luminosity (using the \textsc{MYTorus} scattered AGN model for 3C 223) and unfilled data points denote the observed 2-10 keV luminosity. The $L_{\rm 2-10keV} - L_{\rm [O\,III,intrinsic]}$ relation for Type 1 AGN from \citet{panessa2006} is overplotted. Though there is a wide spread in the \citet{panessa2006} relation, most Type 2 quasars have intrinsic X-ray luminosities systematically below the Type 1 relation. Type 2 quasars may be inherently weaker in X-rays, even when accounting for absorption, than their Type 1 counterparts.}
  \end{center}
\end{figure}

In Figure \ref{lo3_v_lx}, we compare the [\ion{O}{3]} and 2-10 keV luminosity of these quasars with the \citet{panessa2006} $L_{\rm 2-10keV}$ vs. $L_{\rm [O\,III]}$ relation for Type 1 Seyferts and quasars: Log($L_{\rm 2-10 keV}$) = ($0.95\pm0.07$)$L_{\rm [O\,III]}$ + (3.87 $\pm$ 2.76). Their Sy1 sample was selected from the \citet{ho1997} catalog and supplemented with quasars from \citet{mulchaey} and \citet{alonso-herrero}. For this analysis, \citet{panessa2006} used [\ion{O}{3}] fluxes reported in \citet{ho1997} that were corrected for reddening within the NLR. We followed this prescription to estimate the intrinsic [\ion{O}{3}] luminosity ($L_{\rm [O\,III],intrinsic}$) using the observed Balmer decrement (i.e., H$\alpha$/H$\beta$) normalized by an intrinsic value of 3.1 \citep[see][]{osterbrock2006} and the extinction law of \citet{cardelli}:
\begin{equation}
  L_{\rm[O\,III],intrinsic} = L_{\rm [O\,III],observed}\,\left(\frac{L_{\rm H\alpha}/L_{\rm H\beta}}{3.1}\right)^{3.2},
\end{equation}
(see Appendix \ref{o3corr_derivation} for the derivation). With an observed Balmer decrement of 4.15, $L_{\rm [O\,III],intrinsic} = 5.81\times10^{42}$ erg s$^{-1}$ for 3C 223. Due to the high redshift of some of the Type 2 SDSS-selected quasars in the \citet{lansbury2014,lansbury2015} samples, H$\alpha$ does not fall within the observed optical spectrum, so the [\ion{O}{3}] luminosities for these sources are lower limits in Figure \ref{lo3_v_lx}. For all three BASS Type 2 quasars, the observed Balmer decrement was lower than the assumed intrinsic value, so no reddening correction was applied.

Figure \ref{lo3_v_lx} shows that nearly all the Type 2 quasars considered here have X-ray luminosities below the $L_{\rm 2-10 keV} - L_{\rm [O\,III]}$ relation from \citet{panessa2006}. Though there is a wide spread on the best-fit values for this relation, the suppressed X-ray emission indicates that even after correcting for absorption, Type 2 quasars are more X-ray weak than their Type 1 counterparts.

\subsection{The Dearth of Compton-Thick Radio Loud AGN?}\label{cthick_dearth}
\citet{ursini} posed an interesting question: whither Compton-thick radio loud AGN? They claimed that when obtaining high quality hard X-ray spectra ($>$10 keV) of Compton-thick radio loud AGN candidates, there has yet to be unequivocal evidence of column densities exceeding $1.25\times10^{24}$ cm$^{-2}$ measured from models that accurately treat the effects of Compton scattering (e.g., \textsc{MYTorus}, \citealp{mytorus}; \textsc{sphere} and \textsc{torus}, \citealp{brightman2011}; \textsc{borus02}, \citealp{balokovic}).They analyzed archival \textit{NuSTAR} spectra of three radio loud, Type 2 heavily absorbed AGN candidates from \citet{panessa2016} using \textsc{MYTorus} in the default ``coupled'' configuration, freezing the inclination angle at 90$^{\circ}$ (NGC 612, 4C 73.08, 3C 452). The fitted equatorial column densities range from $\sim 4 - 9 \times 10^{23}$ cm$^{-2}$.

To date, there have been two other radio loud, Type 2 AGN observed by {\it NuSTAR} and modeled with \textsc{MYTorus} that allow for a direct comparison with our results. Centaurus A has a fitted equatorial column density of 1.10$^{+0.15}_{-0.02} \times 10^{23}$ cm$^{-2}$ for an inclination angle $\geq 76^{\circ}$ \citep{fuerst}, which is consistent with its line-of-sight column density of $\sim1.1\times10^{23}$ cm$^{-2}$ measured from {\it Suzaku} observations \citep{tzanavaris2021}. Cygnus A, whose X-ray spectra includes contributions from the intracluster medium in which it is embedded and shows signatures of a fast, highly ionized outflow, is obscured by a line-of-sight column density of $\sim 1.6 \times 10^{23}$ cm$^{-2}$. However, when \citet{reynolds} fit the X-ray spectra with \textsc{MYTorus} in decoupled mode, they find an average global column density that is Compton-thick ($N_{\rm H,global} = 1.9^{+1.0}_{-0.8} \times 10^{24}$ cm$^{-2}$).

We find a comparable column density for 3C 223 to those reported in the literature for other Type 2, radio loud AGN observed with {\it NuSTAR} ($1-9 \times 10^{23}$ cm$^{-2}$): heavily obscured, but not Compton-thick.\footnote{For \textsc{MYTorus} inclination angles of 90$^{\circ}$, the equatorial column density ($N_{\rm H,equatorial}$) from the ``coupled'' configuration is equivalent to the line-of-sight column density ($N_{\rm H,los}$) in our ``decoupled'' configuration. Thus our $N_{\rm H,los}$ measurement can be directly compared to the equatorial $N_{\rm H}$ values for NGC 612, 4C 73.08, and 3C 452 reported in \citet{ursini}.} Whether this heavy obscuration is along the line of sight or represents the global average column density depends on the spectral model. 

Interestingly, clear cases of globally Compton-thick radio loud AGN comes from two Type 1 (broad line) AGN: Mrk 668 \citep{sobolewska} and 4C 74.26 \citep{tzanavaris2019}. Mrk 668, a Gigahertz Peaked-Spectrum source, was first suggested to be Compton-thick based on features of reprocessing in its {\it XMM-Newton} spectrum \citep{guianazzi2004}. Follow-up observations with {\it NuSTAR} and modeling of the {\it Chandra} plus {\it NuSTAR} spectra with the \textsc{MYTorus} and \citet{balokovic} \textsc{torus} model demonstrated that the torus is patchy with a global column density that is Compton-thick and about four times higher than the (Compton-thin) line-of-sight column density \citep{sobolewska}. Similarly, the {\it Suzaku} and {\it NuSTAR} spectra of 4C 74.26 were fitted with \textsc{MYTorus} to reveal a patchy obscuring medium that is Compton-thick globally while Compon-thin along the line of sight \citep{tzanavaris2019}.

Radio loud AGN enveloped in Compton-thick levels of obscuring gas do exist. Contrary to expectations from simple one-dimensional models, this Compton-thick material is not observed along the line of sight but is present in the global average column density, and imprints signatures of Compton scattering on the observed X-ray spectrum. Currently, there are three known examples of globally Compton-thick radio loud AGN, one of which is optically obscured (Cygnus A) and two of which are optically unobscured (Mrk 668 and 4C 74.26). Additionally, there are instances of heavily obscured radio loud AGN that have complex environments that are best fit by absorption models requiring distinct column densities \citep[e.g., 4C+29.30,][]{sobolewska2012}. More high-energy X-ray observations above 10 keV are needed to test whether the fraction of radio loud AGN with Compton-thick reprocessing gas is systematically lower than radio quiet AGN, and whether such a trend, if it does exist, implicates evolutionary or environmental origins.

\section{Conclusions}

We analyzed the {\it XMM-Newton} and {\it NuSTAR} spectra of 3C 223, a radio loud Type 2 quasar using the physically-motivated, self-consistent \textsc{borus02} model \citep{balokovic} and \textsc{MYTorus} model \citep{mytorus}, testing both a scattered AGN origin (``\textsc{MYTorus} scattered AGN'') and jet origin (``\textsc{MYTorus} jet'') for the soft emission. We find that the broad band X-ray spectra (0.5 - 30 keV) are best described by a model where the X-ray reprocessing material is patchy, consisting of gas clouds with a high, though not Compton-thick, column density ($N_{\rm H} > 10^{23}$ cm$^{-2}$), and gas clouds with much lower column densities (several $\times 10^{22}$ cm$^{-2}$). The spectral models give conflicting results about whether the high column density gas is distributed along the line of sight (\textsc{MYTorus} scattered AGN model), or represents the global average column density (\textsc{MYTorus} jet model and \textsc{borus02} model; Tables \ref{mytor_decoup_freeas_table}-\ref{borus_table}). The inhomogeneous nature of the X-ray obscuring medium confirms the results presented in \citet{lamassa2014} from analysis of the {\it XMM-Newton} data only. Though all three models provide good fits to the spectra of similar statistical significance, the latter two models find a best-fit AGN spectral slope at the lower limit allowed by the models ($\Gamma_{\rm limit}$ = 1.4), while the spectral slope from the \textsc{MYTorus} scattered AGN model is consistent with typical AGN values \citep[$\Gamma$ = 1.91$^{+0.33}_{-0.27}$;][]{reeves2000}. The \textsc{borus02} model fit gives an elevated iron abundance that is about twice that of the solar value.

We find a relatively high normalization between between the transmitted continuum and the Compton scattered emission ($A_{\rm s} \sim 5-6$) for the \textsc{MYTorus} models and an unphysically high scattering fraction ($f_{\rm scatt} \sim 27\%$) for the \textsc{borus02} model, even when allowing the iron abundance to be a free parameter in the spectral fitting. These fitted parameters indicate that additional physics beyond those accounted for in the \textsc{MYTorus} and \textsc{borus02} calculations are shaping the emergent spectrum, namely the strength of the Fe K$\alpha$ line compared with the continuum. Some possibilities, illustrated in Figure \ref{geometry-sketches}, include:
\begin{itemize}
\item {\it Variability}: Since no variability was observed in the 18 year window between {\it XMM-Newton} epochs (Appendix \ref{xray_var}), this scenario would require the Compton-scattering region to extend beyond 5.5 pc (i.e., the 18 year light-travel time between {\it XMM-Newton} observations). Detailed studies of local AGN demonstrate that most of the Fe K$\alpha$ line emission originates on nuclear scales \citep[e.g.,][]{gandhi}, but there is an extended Comtpon-scattering region beyond the several-parsec torus \citep{bauer2015,fabbiano2017,fabbiano2018,jones2020,yi2021}.
\item {\it Inner Compton-thick ring}: A ring of heavily Compton-thick material with a column density exceeding 10$^{26}$ cm$^{-2}$ could completely block $\sim$80\% of the light in the equatorial direction, causing the global distribution of clouds to see about five - six times more radiation than clouds along the line of sight. We note that modeling the spectra with \textsc{UxClumpy} \citep{buchner2019} to test for the presence of a $N_{\rm H} > 10^{25}$ cm$^{-2}$ ring did not provide an acceptable fit (Appendix \ref{uxclumpy_fitting}).
\item {\it Interaction of jet with the torus}: The jet may be imparting a great deal of energy beamed in a direction not along the line of sight, enhancing Fe K$\alpha$ emission in the global medium relative to the line of sight.
\end{itemize}

With future {\it ATHENA} X-ray Integral Field Unit observations \citep{barret}, which will have $\sim$5$^{\prime\prime}$ pixels and a spectral resolution of 2.5 eV, we could test how much of the Fe K$\alpha$ emission originates from extended scales (5$^{\prime\prime} \sim$ 10 kpc for 3C 223) and whether it is aligned with the more distant radio and X-ray lobes which would help distinguish between the scenarios above.

From the intrinsic (absorption-corrected) X-ray luminosity and [\ion{O}{3}] luminosity, we estimate a radiative bolometric luminosity of $\sim 1-6\times 10^{45}$ erg s$^{-1}$ and an Eddington ratio of $\gtrsim$5\%, consistent with radiatively efficient accretion (Section \ref{edd}). The bolometric luminosity, and associated Eddington ratio, calculated from the [\ion{O}{3}] line is greater than that calculated from the X-ray luminosity, but it is unclear whether this result is due to the radio jet boosting the [\ion{O}{3}] line luminosity \citep{vanbreugel, siemiginowska, couto} or whether an inner, heavily Compton-thick ring of obscuration is causing the intrinsic X-ray luminosity deduced by the \textsc{MYTorus} model to be under-predicted. The amount of kinetic energy carried by the jet  ($2.1 \times 10^{45}$ erg s$^{-1}$) is comparable to that of the radiatively bolometric luminosity (Section \ref{radio}). 3C 223 and other Type 2 [\ion{O}{3}]-defined quasars observed by {\it NuSTAR} generally have suppressed X-ray emission compared with Type 1 AGN studied by \citet[][Figure \ref{lo3_v_lx}]{panessa2006}, suggesting that Type 2 quasars may be inherently weaker in X-rays than their Type 1 counterparts.

Finally, we return to the question of whether there is a dearth of radio loud Compton-thick AGN. From reviewing the literature, we found three radio loud AGN observed by {\it NuSTAR} whose X-ray spectra are well fit by models that measure a Compton-thick global average column density, yet Compton-thin line-of-sight column density: Type 2 (optically obscured) AGN Cygnus A \citep{reynolds} and Type 1 (optically {\it unobscured}) AGN Mrk 668 \citep{sobolewska} and 4C 74.26 \citep{tzanavaris2019}. Current evidence demonstrates that globally Compton-thick radio loud AGN do exist. However, these confirmed cases have complicated spectra that require physically motivated models to disentangle the global from the line-of-sight column density. The presence of Compton-thick gas in these sources could be missed by simpler one dimensional models or those that assume a homogeneous obscuring medium. As the number of detailed broadband X-ray spectroscopic studies of individual AGN increases \citep{lamassa2014,yaqoob2015,balokovic,reynolds,sobolewska,tzanavaris2021}, so too does the census of those that have demonstrable evidence that the X-ray reprocessor is non-uniform, supporting an emerging picture that the gaseous environment that reprocesses emission around an accreting black hole can be complex and that binary characterizations of Compton-thin and Compton-thick oversimplifies reality.

\begin{acknowledgements}

We thank the anonymous referee for a careful reading of this manuscript and thoughtful comments which improved the quality of the paper. S.M.L acknowledges support from NASA grants 80NSSC20K0261 and 80NSSC20K0837. S.M.L is thankful for the efforts of STScI IT, facilities, and administrative staff that allowed continual operations at STScI, even during difficult conditions in the early stages of the COVID-19 pandemic. This paper would not be possible without their support. A.S. was supported by NASA contract NAS8-03060 (Chandra X-ray Center).

Based on observations obtained with XMM-Newton, an ESA science mission with instruments and contributions directly funded by ESA Member States and NASA. This research has made use of the NuSTAR Data Analysis Software (NuSTARDAS) jointly developed by the ASI Space Science Data Center (SSDC, Italy) and the California Institute of Technology (Caltech, USA). This research has made use of the NASA/IPAC Extragalactic Database (NED), which is funded by the National Aeronautics and Space Administration and operated by the California Institute of Technology. The scientific results reported in this article are based in part on data obtaianed from the Chandra Data Archive. This research has made use of software provided by the Chandra X-ray Center (CXC) in the application package CIAO. This research has made use of the NASA/IPAC Extragalactic Database (NED), which is funded by the National Aeronautics and Space Administration and operated by the California Institute of Technology.

\end{acknowledgements}

\facilities{XMM,NuSTAR,CXO}

\software{XSpec (v12.11.1; Arnaud 1996), NuSTARDAS (v2.0.0), CIAO (v4.13; Fruscione et al. 2006)}

\object{3C 223}

\clearpage

\appendix

\section{Variability}\label{xray_var}
We investigated whether the X-ray spectrum of 3C 223 exhibits evidence of variability between the initial {\it XMM-Newton} observations from 2001 and the most recent observations from 2019. For this test, we included analysis of a short (8 ks) {\it Chandra} observation of 3C 223 taken in 2012 as part of a snapshot survey of 3C radio sources \citep[PI: Harris, ObsID: 12731;][]{massaro2012}. The {\it Chandra} data were processed with \textsc{ciao} v. 4.13 and CALDB 4.9.5 \citep{fruscione}. We ran the recommended \textsc{chandra\_repro} script to generate a calibrated events file from which we extract the {\it Chandra} spectrum of 3C 223, with the source spectra extracted from a 3$^{\prime\prime}$ radius circle centered on the peak X-ray emission of the source, and the background derived from an annulus with an inner radius of 5$^{\prime\prime}$ and 20$^{\prime\prime}$. We note that in this short exposure, we see no evidence of the extended X-ray lobes reported in the {\it XMM-Newton} spectra by \citet{croston} and expect that this region is dominated by X-ray background photons. The spectra were binned with {\textsc ftgrouppha}, with a minimum S/N of 2 per bin in the background subtracted spectrum.

\begin{figure*}[h]
  \begin{center}
    \includegraphics[scale=0.38]{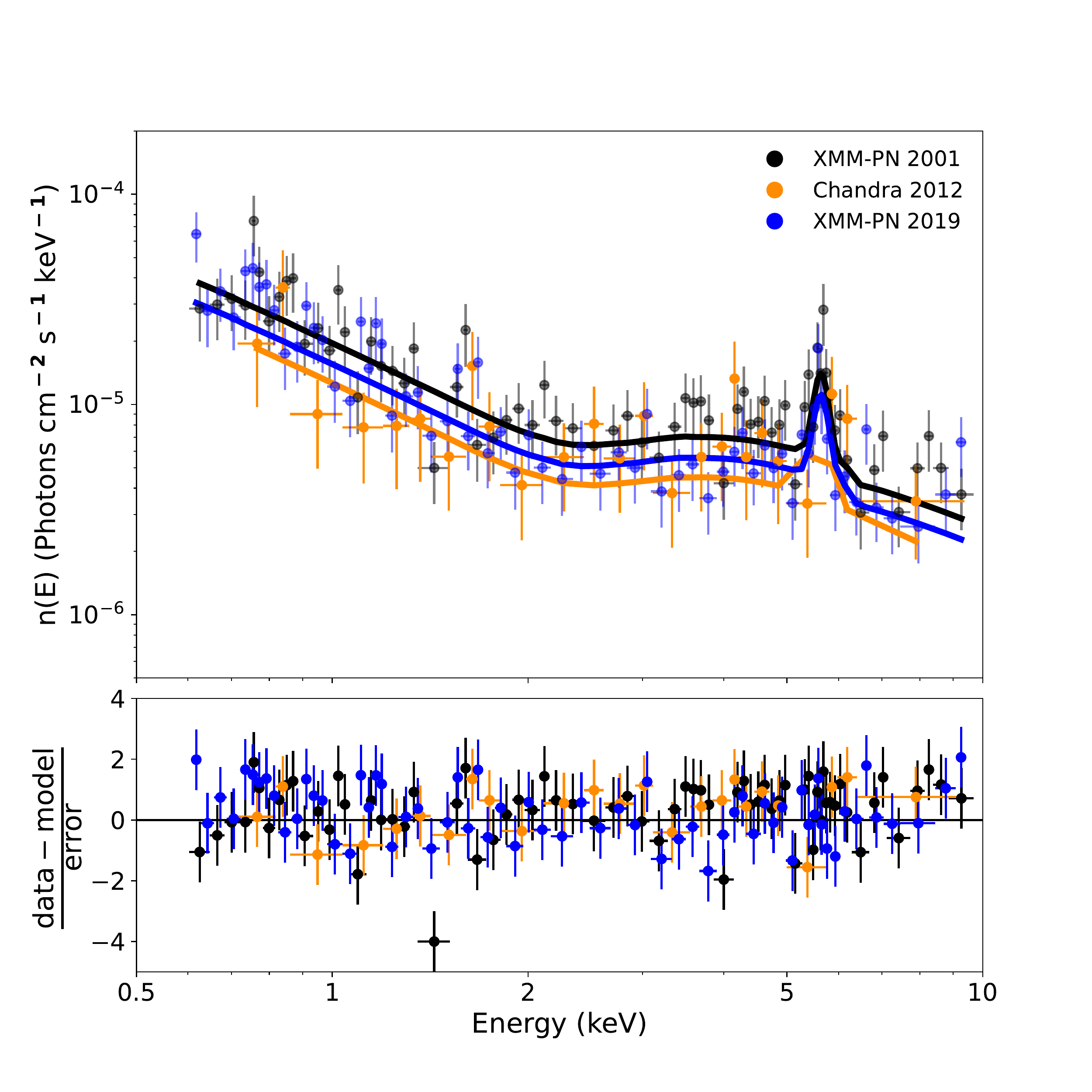}
\caption{\label{chandra_xmm_pow} Results from fitting the {\it XMM-Newton} and {\it Chandra} spectra with a phenomenological double powerlaw + Gaussian model to search for indications of variability between epochs. No deviations in the observed spectra from the best-fit model are evident, indicating that the intrinsic X-ray spectrum has not varied significantly at the times of these observations. The overall normalization for the {\it Chandra} spectrum is lower than that of the {\it XMM-Newton} spectra, which could be instrumental or due to unresolved X-ray emission being captured in the deeper {\it XMM-Newton} observations. For clarity, only the spectra from the {\it XMM} PN detector are plotted, though we included the MOS1 and MOS2 spectra from both epochs when fitting the dataset.}
  \end{center}
\end{figure*}

We fitted the two epochs of {\it XMM-Newton} spectra and the {\it Chandra} spectrum between 0.5-10 keV with a phenomenological double absorbed powerlaw model with a Gaussian component for modeling the Fe K$\alpha$ emission. A constant multiplicative factor was included in the modeling to serve as the cross-detector and cross-instrument normalization. Despite concerted cross-calibration efforts across missions, systematic differences up to 10-15\% in the absolute flux calibration often remain \citep[e.g.,][]{madsen2017}. The results of this spectral fitting are shown in Figure \ref{chandra_xmm_pow}, where we show only the {\it XMM-Newton} PN spectra from 2001 and 2019 for clarity, though the MOS1 and MOS2 spectra for both epochs were included in the fitting.

The intrinsic spectrum shows no evidence of variability from 2001 to 2012 to 2019: none of the spectra show systematic deviations from the best-fit model. The overall normalization between spectral epochs show some differences, The average normalization of the {\it XMM-Newton} spectra from 2001 (2019) is 1.00$\pm$0.18 (0.82$\pm$0.17), with the {\it Chandra} normalization equal to 0.64$^{+0.12}_{-0.11}$. These cross-calibration normalizations are consistent within the 90\% confidence interval between the {\it XMM-Newton} spectra and between the {\it Chandra} spectrum and {\it XMM-Newton} spectra from 2019. The systematically lower normalization for the {\it Chandra} spectrum, though not statistically significant, could be due to unresolved emission at scales larger than the $3^{\prime\prime}$ radius circular aperture used for the {\it Chandra} spectrum that is observed in the deeper {\it XMM-Newton} data extracted from the larger 30$^{\prime\prime}$ radius aperture.

\section{Ruling out other MYTorus model fits}\label{ruled_out_fits}
\begin{figure*}[h]
  \begin{center}
    \includegraphics[scale=0.38]{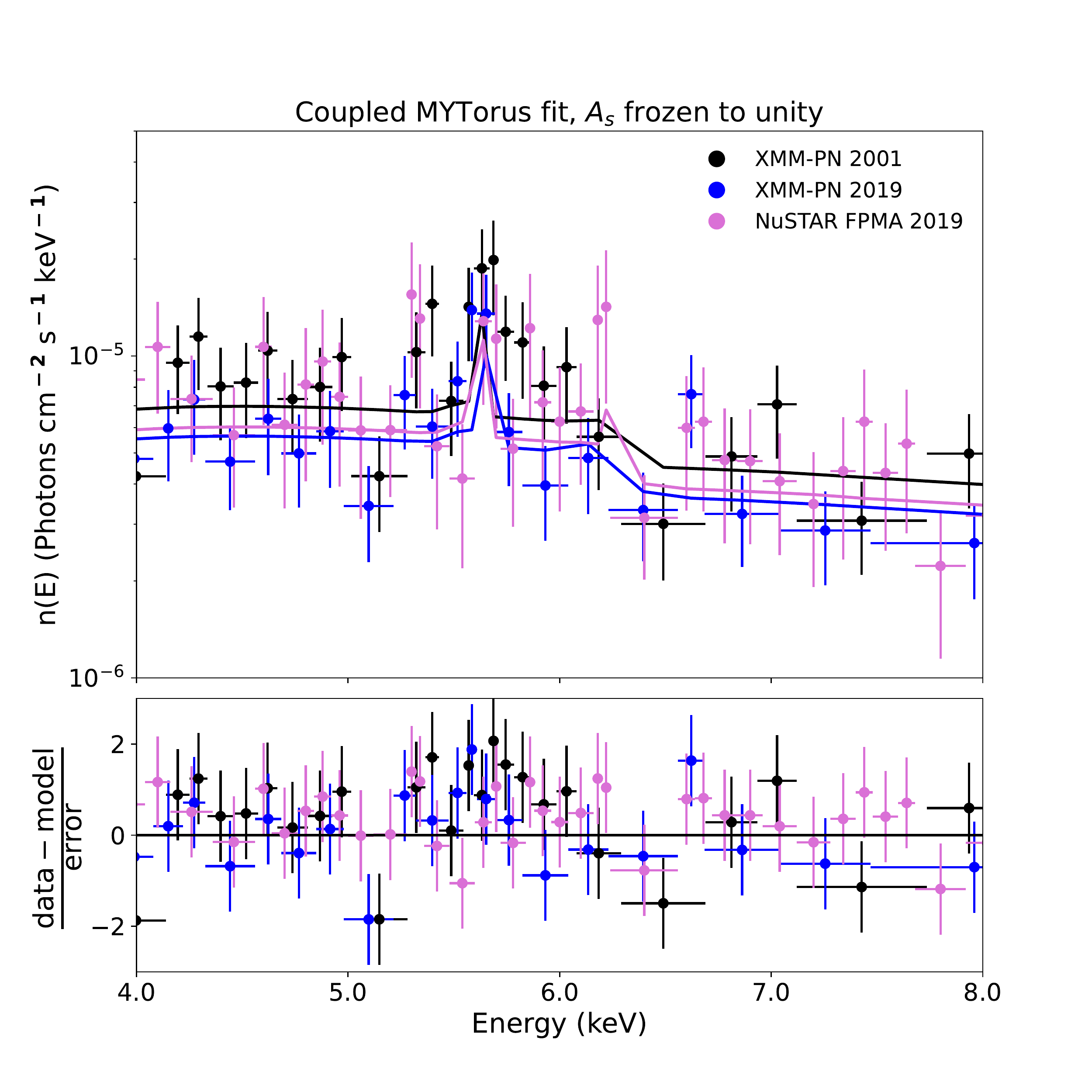}~
    \hspace{-0.8cm}
  \includegraphics[scale=0.38]{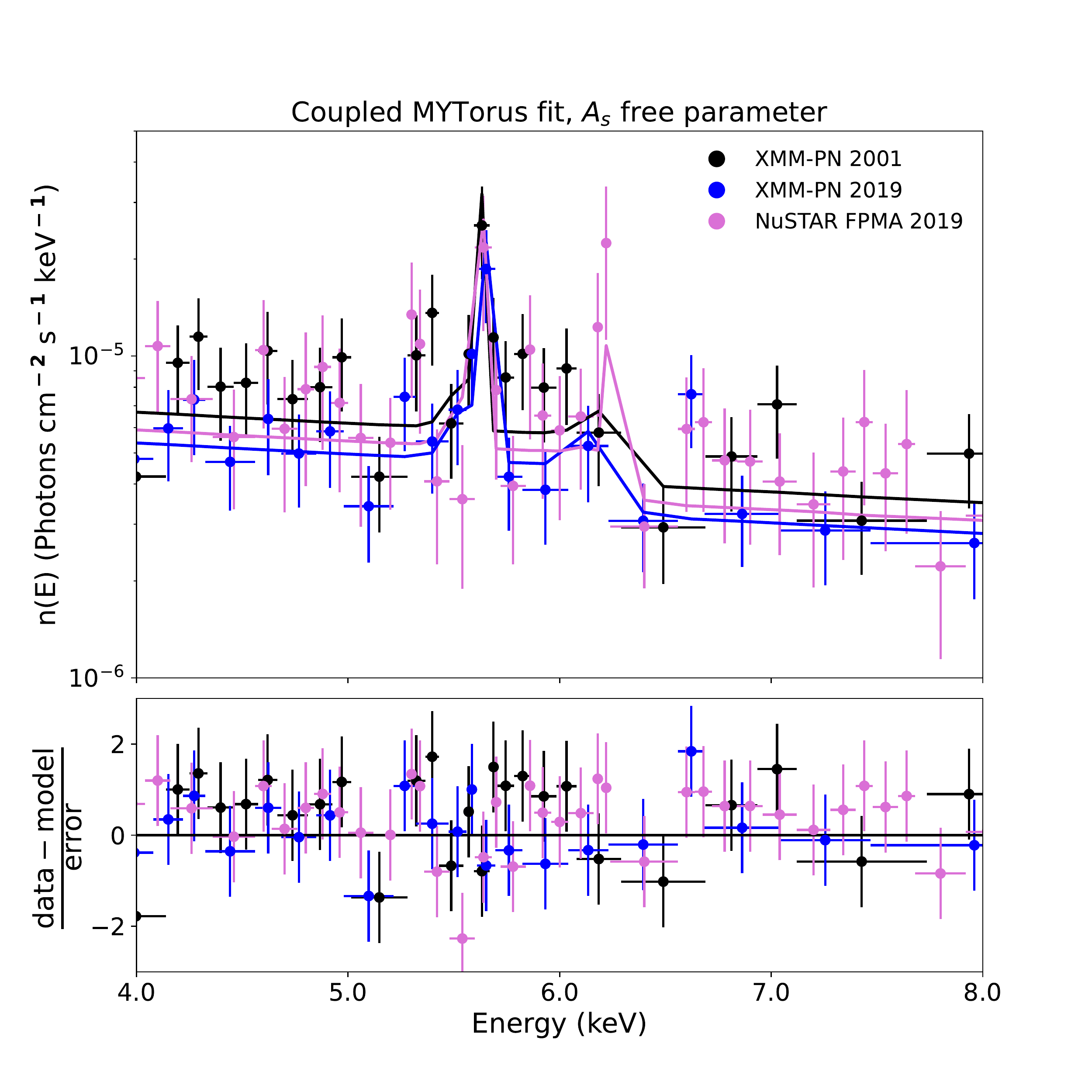}\\
  \includegraphics[scale=0.38]{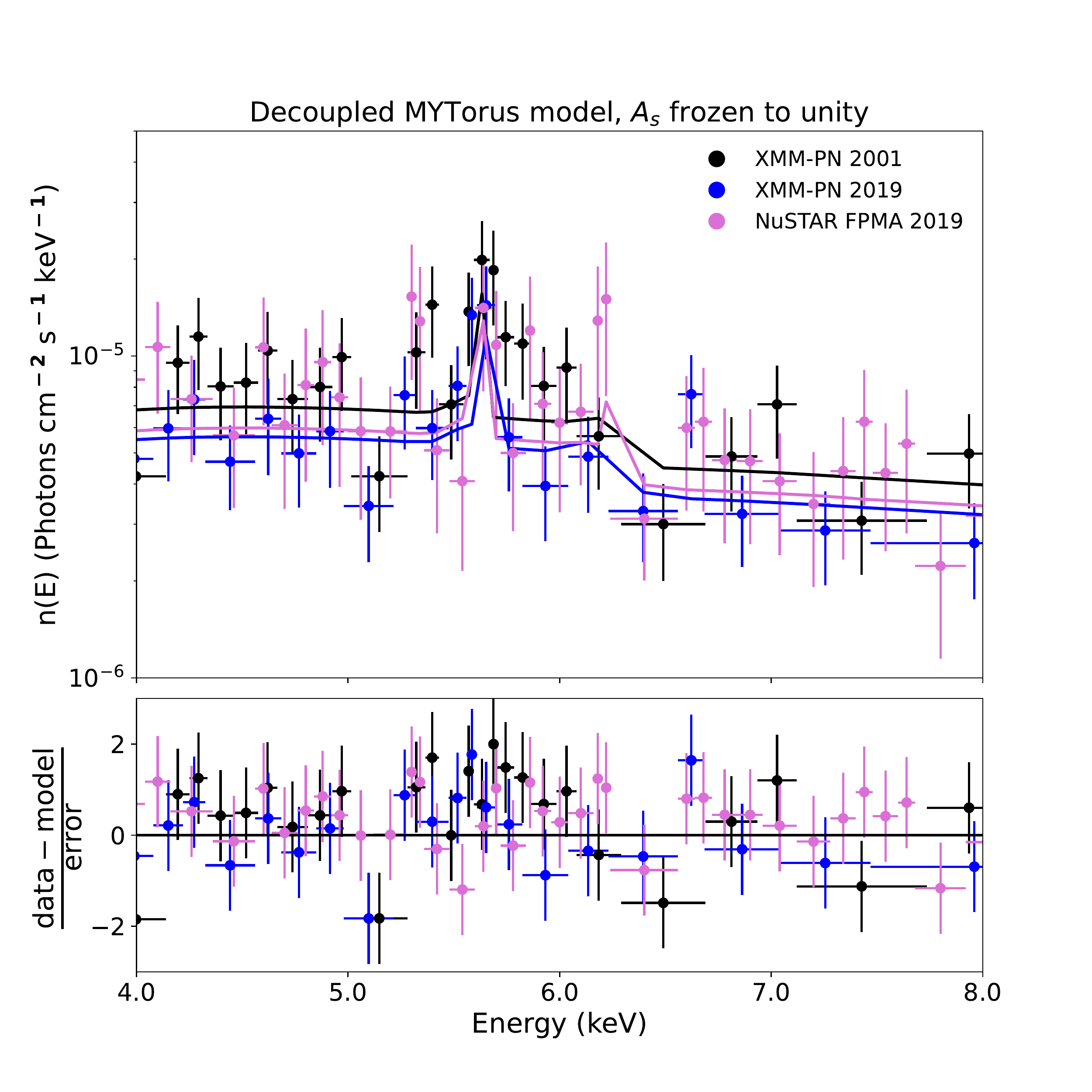}~
  \hspace{-0.8cm}
  \includegraphics[scale=0.38]{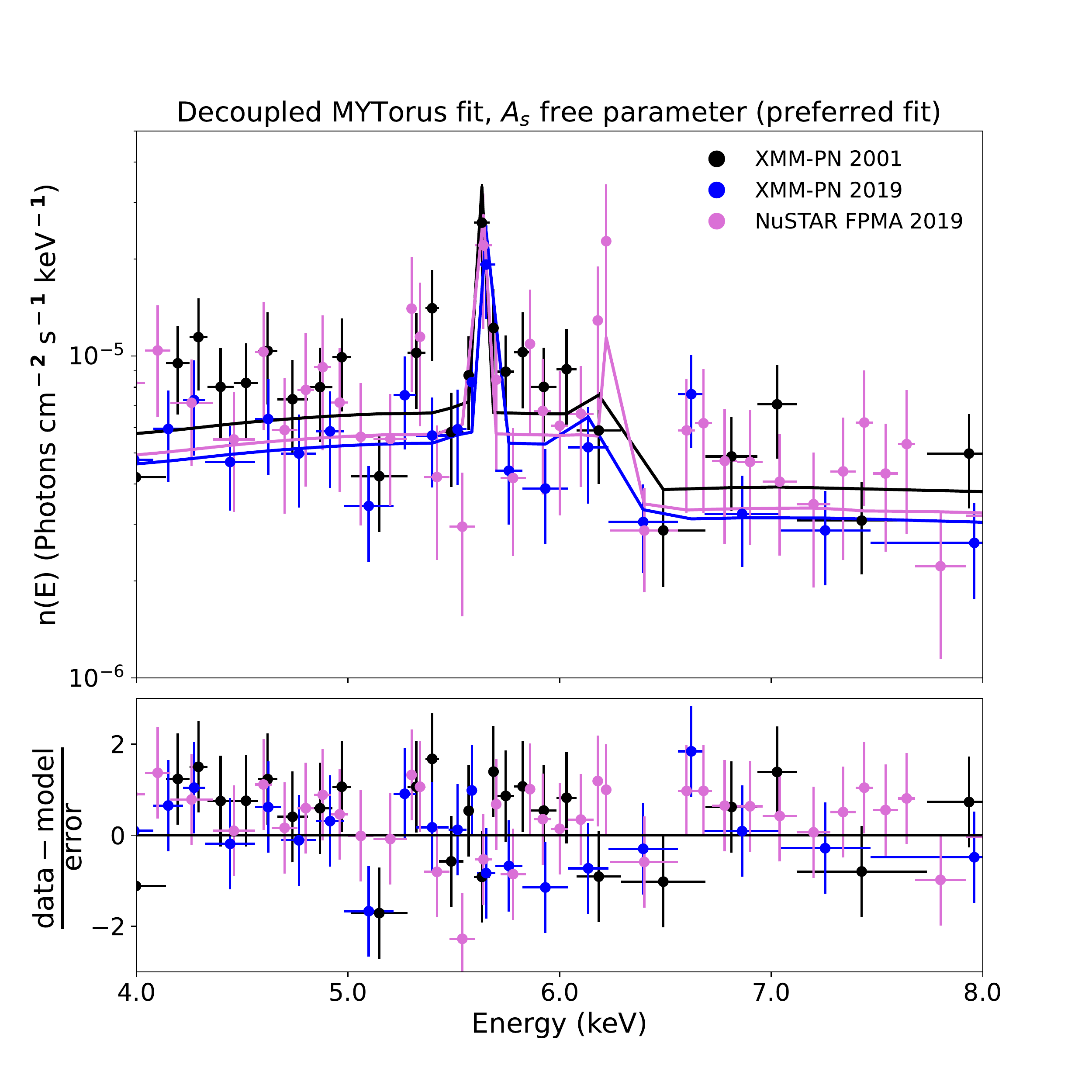}
  \caption{\label{feka_plots} Close-up of the fitted Fe K$\alpha$ complex for various \textsc{MYTorus} model realizations.  Both the coupled and decoupled \textsc{MYTorus} model fits where $A_{\rm S}$ was frozen at unity (left panels) exhibit significant residuals around the Fe K$\alpha$ line and local continuum. Though the coupled \textsc{MYTorus} model where $A_{\rm S}$ was a free parameter (top right) shows a much better fit around the Fe K$\alpha$ complex, we reject this model fit due to the fitted grazing incidence angle of the torus (see Table \ref{rej_mytor_table_scatt}). The fit to the Fe K$\alpha$ line and local continuum for our preferred model fit (decoupled \textsc{MYTorus} scattered AGN model with $A_{\rm S}$ fitted as a free parameter) is shown in the bottom right for reference. Only a subset of the data are plotted for ease of visualization, though all spectra were fitted.}
  \end{center}
\end{figure*}

When trying multiple \textsc{MYTorus} configurations to fit the X-ray spectra of 3C 223, we modeled the soft X-ray emission (0.5 - 2 keV) as originating from scattered AGN light or from the base of a radio jet. We summarize the results in Table \ref{rej_mytor_table_scatt} and Table \ref{rej_mytor_table_jet}, ascribing the soft emission to scattered AGN flux or jet emission, respectively. In Figure \ref{feka_plots}, we show close ups around the Fe K$\alpha$ line for the scattered AGN model only since the plots for the jet model look identical, with the exception of the decoupled \textsc{MYTorus} jet model with a frozen $A_{S}$ value at unity, which shows a similar fit to Fe K$\alpha$ as the decoupled \textsc{MYtorus} jet model where $A_{S}$ is a free parameter. Nevertheless, we reject the former model description as allowing $A_{S}$ to be a free parameter improves the fit at a significant level according to the $F$-test ($F$ statistical value of 16.03 with a probability of 7.4$\times10^{-5}$ that the improvement is due to chance).

In the coupled \textsc{MYTorus} realization, the torus is assumed to be azimuthally symmetric and homogeneous and the fitted column density respresents the equatorial column density ($N_{\rm H,equatorial}$). In these cases, the fitted inclination angle of the torus is constrained to be at the boundary of intersecting the line of sight ($\sim$60$^{\circ}$). Such ``grazing'' inclination angles indicate that the model is reconciling the need for a high global column density to account for strong Fe K$\alpha$ emission relative to the continuum (i.e., high Fe K$\alpha$ EW) with an observed continuum shape that does not have the severe low-energy spectral curvature that would be associated with such high-to-Compton-thick column densities. Fitted grazing inclination angles are thus a clue that the X-ray reprocessor is not homogeneous.

We also explored \textsc{MYTorus} model options where the relative normalization between the transmitted and Compton scattered components of the model ($A_{\rm S}$) is frozen to unity and when it is allowed to be fitted as a free parameter. We found that when $A_{\rm S}$ is constrained to unity, both the Fe K$\alpha$ line and continuum around the Fe K$\alpha$ line is poorly fit (Figure \ref{feka_plots}, left panel), compared to our best fit model (Figure \ref{feka_plots}, bottom right panel). Though the coupled \textsc{MYTorus} model fit with free $A_{\rm S}$ shows an acceptable fit around the Fe K$\alpha$ complex (Figure \ref{feka_plots}, upper right), this model can be ruled out due to the grazing inclination angle found for the torus that disfavors a homogenous obscuring medium as assumed by the default \textsc{MYTorus} model set-up.

\begin{deluxetable}{llll}[h]
\tablecaption{\label{rej_mytor_table_scatt}\textsc{MYTorus} Fit Parameters for Rejected Models: Scattered AGN Emission\tablenotemark{a}}
\tablehead{
  \colhead{Parameter} & \multicolumn{2}{c}{Coupled} &  \colhead{Decoupled} \\
}
\startdata
$\Gamma$                                   & 1.55$^{+0.05}_{-0.03}$ & 1.69$^{+0.04}_{-0.08}$ & 1.55$^{+0.11}_{-0.12}$\\
Power law norm (10$^{-4}$)                 & 1.43$^{+0.17}_{-0.08}$ & 0.93$^{+0.34}_{-0.08}$ & 1.39$^{+0.52}_{-0.39}$ \\
Inclination angle ($^{\circ}$)             & 61.0$^{+3.5}_{-0.6}$ & 60.5$^{+0.5}_{-0.2}$ & \nodata \\
$N_{\rm H,equatorial}$ (10$^{24}$ cm$^{-2}$)  & 0.90$^{+0.64}_{-0.47}$ & 0.96$^{+0.16}_{-0.26}$ & \nodata \\
$N_{\rm H,los}$ (10$^{24}$ cm$^{-2}$)    & \nodata               & \nodata  & 0.21$^{+0.04}_{-0.04}$ \\
$N_{\rm H,global}$ (10$^{24}$ cm$^{-2}$) & \nodata               & \nodata   & 1.00$^{+1.20}_{-0.48}$ \\
$A_{\rm S}$\tablenotemark{b}           & 1 (f)   & 7.9$^{+3.0}_{-1.2}$ &  1 (f) \\
$f_{\rm scatt}$ (10$^{-2}$)             & 17$^{+3}_{-1}$ & 27$^{+2}_{-3}$ & 18$^{+6}_{-5}$ \\
$\chi^2$ (dof)                        & 364.0 (396) & 338.6 (395) & 360.0 (396) \\
\enddata
\tablenotetext{a}{In the coupled realization of the \textsc{MYTorus} model, the X-ray obscurer is assumed to be homogenous, with a fixed covering fraction of 0.5. The measured column density denotes the equatorial column density ($N_{\rm H,equatorial}$). This table reports fit parameters where soft X-ray emission (0.5 - 2 keV) is modeled as originating from scattered AGN light.}
\tablenotetext{b}{Normalization between the transmitted AGN continum and the Compton scattered and fluorescent line emission, which was either left frozen at unity or allowed to be free a parameter.}
\end{deluxetable}

\begin{deluxetable}{llll}[h]
\tablecaption{\label{rej_mytor_table_jet}\textsc{MYTorus} Fit Parameters for Rejected Models: Jet Model\tablenotemark{a}}
\tablehead{
  \colhead{Parameter} & \multicolumn{2}{c}{Coupled} &  \colhead{Decoupled} \\
}
\startdata
$\Gamma_{\rm AGN}$                         & $<$1.49 & $<$1.44 & $>$2.44 \\
Power law norm - AGN (10$^{-4}$)          & 1.04$^{+0.10}_{-0.21}$ & 0.53$^{+0.07}_{-0.06}$ & 85.5$^{+5.8}_{-32.8}$ \\
$\Gamma_{\rm Jet}$                         & 1.62$^{+0.15}_{-0.13}$ & 1.86$^{+0.24}_{-0.22}$ & 1.84$^{+0.37}_{-0.27}$ \\
Power law norm - Jet (10$^{-4}$)           & 0.25$^{+0.02}_{-0.01}$ &  0.26$\pm0.02$ & 0.23$\pm0.02$\\
Inclination angle ($^{\circ}$)             & 60.9$^{+2.8}_{-0.5}$ &  60.5$^{+0.5}_{-0.2}$ & \nodata \\
$N_{\rm H,equatorial}$ (10$^{24}$ cm$^{-2}$)  & 0.80$^{+0.50}_{-0.40}$ & 0.60$^{+0.10}_{-0.23}$ & \nodata \\
$N_{\rm H,los}$ (10$^{24}$ cm$^{-2}$)    & \nodata               & \nodata  & 1.11$^{+0.15}_{-0.16}$ \\
$N_{\rm H,global}$ (10$^{24}$ cm$^{-2}$) & \nodata               & \nodata   & 0.19$^{+0.03}_{-0.06}$ \\
$A_{\rm S}$                            & 1 (f)   &  7.2$^{+1.7}_{-2.4}$ &  1 (f) \\
$\chi^2$ (dof)                        & 363.4 (395) & 335.0 (394) & 350.4 (395) \\
\enddata
\tablenotetext{a}{The fit parameters reported in this table are from modeling the soft X-ray emission with a jet origin.}
\tablenotetext{b}{Normalization between the transmitted AGN continum and the Compton scattered and fluorescent line emission, which was either left frozen at unity or allowed to be free a parameter.}
\end{deluxetable}

\section{Fitting \textsc{UxClumpy} to 3C 223 X-ray Spectra}\label{uxclumpy_fitting}
The \textsc{UxClumpy} (Unified X-ray \textsc{Clumpy}) model is based on a geometry where the X-ray reprocessor takes the form of discrete clouds that can have a range of column densities and angular distributions \citep{buchner2019}. \textsc{XSpec} table models are available that are derived from Monte Carlo simulations that use the \textsc{XARS} (X-ray Absorption Re-emission Scattering) code to pre-compute X-ray spectra for various input parameters. The clumpy geometry has the benefit of describing eclipse events observed in X-ray spectra of some AGN \citep{mckernan,risaliti2002,risaliti2005,risaliti2009,risaliti2011,rivers2015,marinucci2016,ricci2016} and interfacing with infrared clumpy torus models \citep{nenkova2008a,nenkova2008b}.

The geometry of the X-ray reprocessor is in part defined by $\sigma_{\rm Torus}$ which sets the angular width of the cloud population: clouds are distributed along the equatorial plane for lower values of $\sigma_{\rm Torus}$ and occupy nearly spherical coverage for the highest $\sigma_{\rm Torus}$ values. $\sigma_{\rm Torus}$ can range from 0$^{\circ}$ (no clumps) to 84$^{\circ}$ (spherical coverage). The fitted column density refers to the line-of-sight column density ($N_{\rm los}$), which similar to the \textsc{MYTorus} model in decoupled mode, measures the column of gas the X-ray photons transverse from the corona to the observer. The \textsc{UxClumpy} model partitions the sky around the X-ray source into bins of torus inclination angle and column density, such that similar inclination angles can have different line-of-sight column densities. $N_{\rm H,los}$ drives the emergent X-ray spectral shape more than the inclination angle.

\textsc{UxClumpy} includes an optional inner uniform Compton-thick ring, where column densities exceed 10$^{25}$ cm$^{-2}$, as a model component. This ring serves as a cold reflecting mirror with a covering factor that can range from 0 (no ring) to 0.6. Some heavily obscured, local AGN require this component to achieve a good fit to their $>$10 keV {\it NuSTAR} spectra (i.e., Circinus, NGC 424, and ESO 103-G035). Physically, such a component might represent a warped accretion disk or the inner wall of the torus \citep{buchner2019}.

\textsc{UxClumpy} self-consistently models soft X-ray emission from scattered AGN light by accounting for the effects of reflection from hot gas (i.e., X-ray photons Compton scattered by hot electrons) and cold dense clouds beyond the putative torus \citep[which can give rise to Fe K$\alpha$ emission and a Compton hump; see e.g.,][]{bauer2015,fabbiano2017,fabbiano2018,jones2020,yi2021}. The \textsc{uxclumpy-scattered} table model represents the angle-averaged spectrum of the warm reflected component whose model parameters are linked to those from the primary \textsc{UxClumpy} model to enforce self-consistency during spectral fitting. \citet{buchner2019} advise that the relative normalization between the transmitted/Compton reflected and warm scattered models should not exceed 10\% which would otherwise give rise to unphysical scattering efficiences. The model is set up in \textsc{XSpec} as:
\begin{eqnarray}
  \begin{array}{ll}
  \mathrm{model} = & const_1 \times phabs \times \nonumber \\
  & (\mathrm{atable\{uxclumpy-cutoff.fits\}} + \nonumber \\
  & const_2 \times \mathrm{atable\{uxclumpy-cutoff-omni.fits\}})
  \end{array}
\end{eqnarray}

\noindent where \textsc{uxclumpy-cutoff.fits} models the transmitted and Compton reflected emission, \textsc{uxclumpy-cutoff-omni.fits} models the warm scattered emission, and \textsc{const\_2} represents the scattering fraction ($f_{\rm scatt}$). Similar to the \textsc{MYTorus} modeling, \textsc{const\_1} parameterizes the cross-instrumental normalization and \textsc{phabs} models the Galactic absorption and is frozen at this value ($1.46 \times 10^{20}$ cm$^{-2}$). We froze the high energy cutoff to the maximum allowed value of 400 keV as our data are not sensitive enough to measure this parameter.

When using this model configuration to fit the X-ray spectra of 3C 223, we initially imposed an upper limit on the scattering fraction of 0.1 (i.e., 10\%), but this resulted in an unconstrained fit on the primary powerlaw continuum (i.e., the error on the powerlaw normalization spanned several orders of magnitude). We thus attempted two model realizations to fit the X-ray spectra of 3C 223 with \textsc{UxClumpy}: freezing the scattering fraction to 0.1 and allowing the scattering fraction to be a free parameter bounded by an upper limit of 1.0 (Table \ref{uxclumpy_table}). We tested for the presence of an inner Compton-thick ring by allowing CTKcover to be a free parameter, but the best-fit covering factor was zero and we were only able to obtain upper limits.

\begin{figure*}[h]
  \begin{center}
    \includegraphics[scale=0.38]{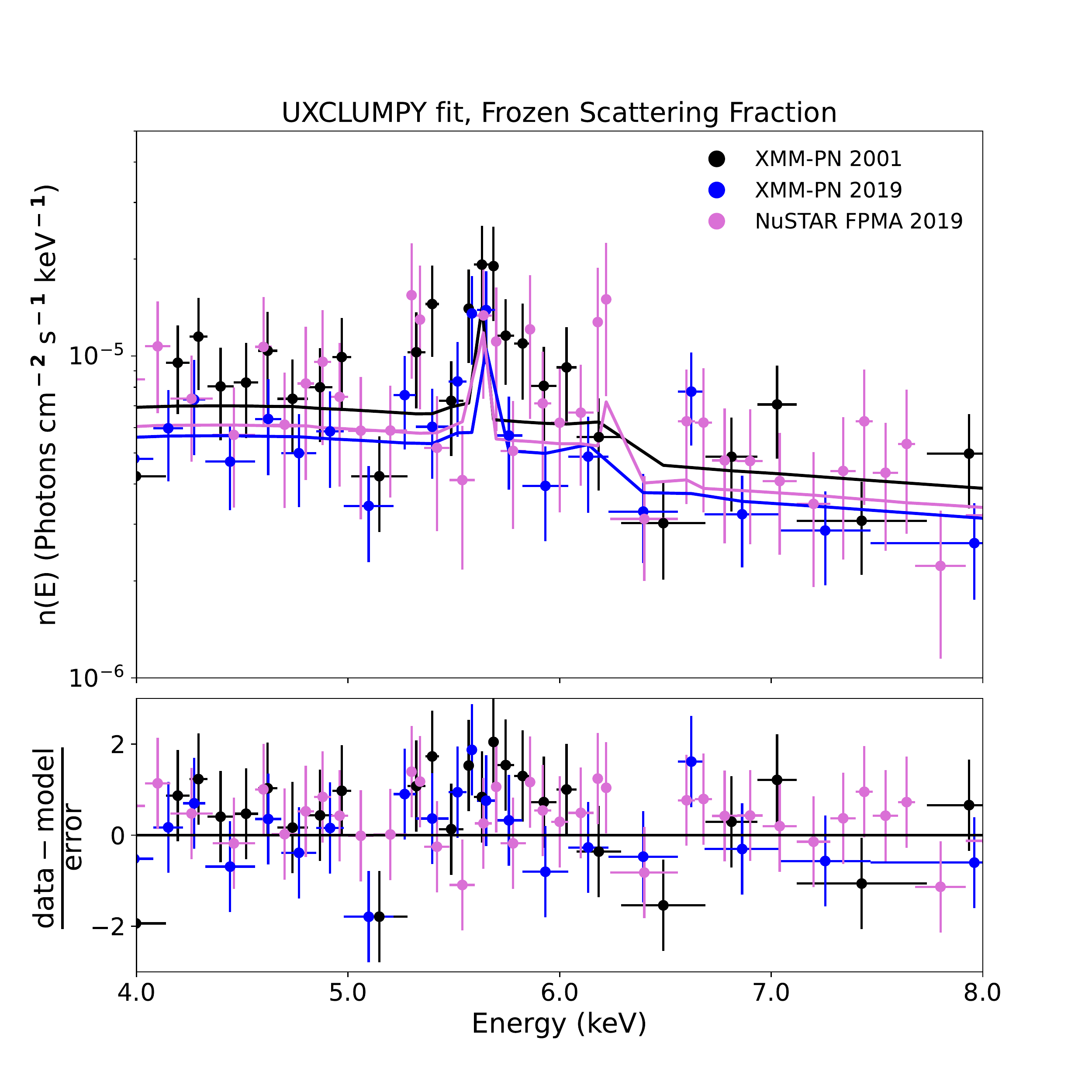}~
    \hspace{-0.8cm}
  \includegraphics[scale=0.38]{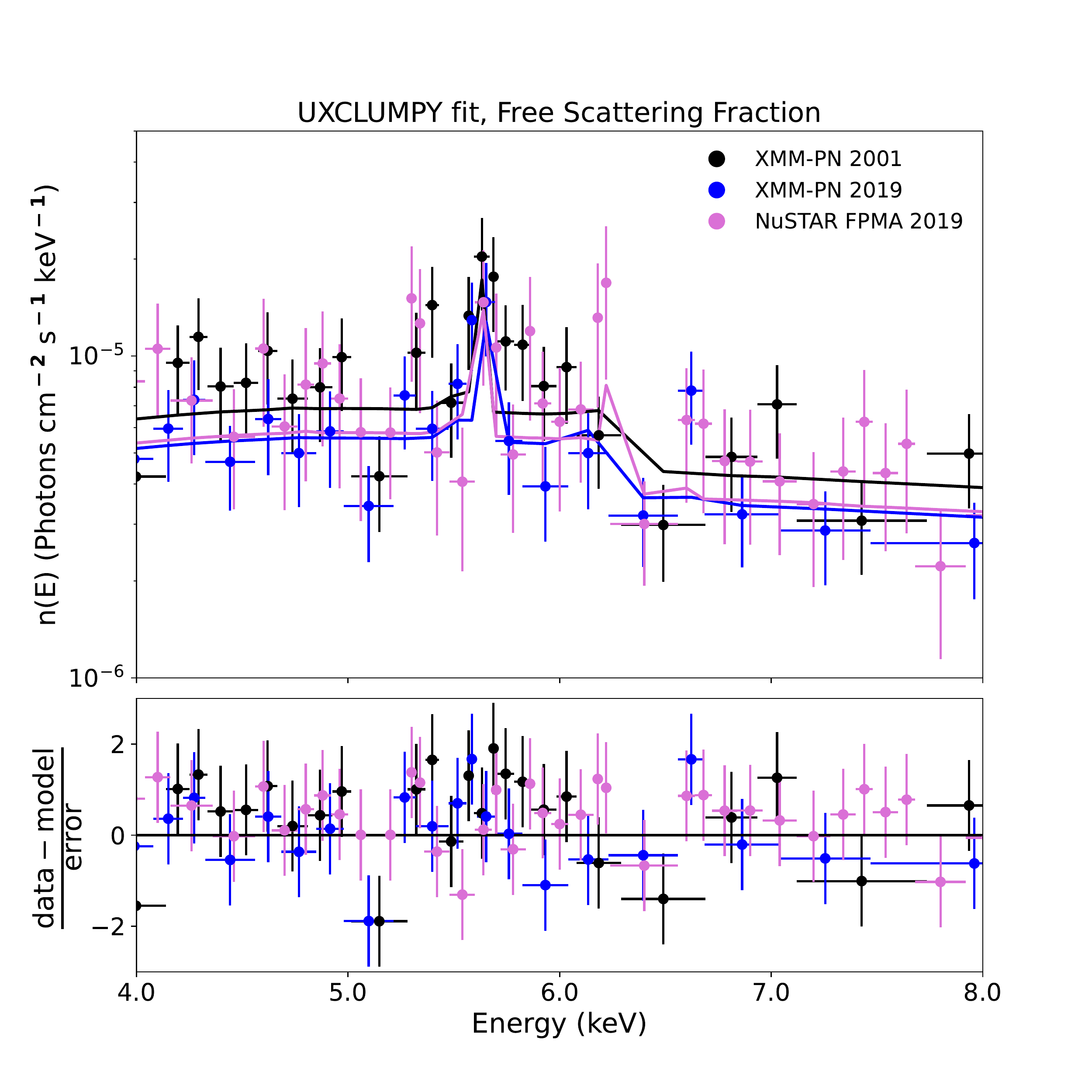}\\
  \caption{\label{feka_plots_uxclumpy} Close-up of the fitted Fe K$\alpha$ complex for \textsc{UxClumpy} model fits of the X-ray spectra of 3C 223 where (\textit{left}) the scattering fraction is frozen to 10\% and (\textit{right}) the scattering fraction is a free parameter (with best-fit $f_{\rm scatt} >$66\%, which gives rise to an unphysically high scattering efficiency). In both cases, the Fe K$\alpha$ line is poorly fit, especially when compared with the accepted model (Figure \ref{feka_plots}, bottom right). For ease of visualization, just the {\it XMM-Newton} PN spectra and the {\it NuSTAR} FPMA spectrum are shown, though all 8 spectra from {\it XMM-Newton} (PN, MOS1, MOS2 from observations in 2001 and 2009) and {\it NuSTAR} (FPMA and FPMB) were fitted simultaneously.}
  \end{center}
\end{figure*}

Neither model realization is an adequate fit to the X-ray spectra of 3C 223. Similar to the rejected \textsc{MYTorus} fits described in Appendix \ref{ruled_out_fits}, the Fe K$\alpha$ line is poorly accommodated by the \textsc{UxClumpy} model (see Figure \ref{feka_plots_uxclumpy}). Allowing the scattering fraction to be a free parameter does improve the fit statistically compared with the fit where $f_{\rm scatt}$ is frozen at 10\%, but the measured scattering fraction is unphysical, with a best-fit value pegged at the imposed limit of 100\%, and the Fe K$\alpha$ line remains poorly fit by this model compared with the preferred model (see Figure \ref{feka_plots}, bottom right). The unphysical scattering fraction likely indicates that the soft emission has contributions beyond the scattering of the primary AGN emission that is not accomodated by the physical set-up of the model. Qualitatively, this result is similar to that of the MYTorus model, where such additional physics is implicated by an elevated normalization factor between the transmitted and reflected component ($A_{\rm S} \sim 5-6$), and the \textsc{borus02} model with an elevated scattering fraction ($f_{\rm scatt} = 27^{+5}_{-3}$\%). Both the \textsc{UxClumpy} model and \textsc{MYTorus} scattered AGN model are consistent in finding heavy line-of-sight obscuration to the central engine ($N_{\rm H} \geq 10^{23}$ cm$^{-2}$).

\begin{deluxetable}{lll}[h]
\tablecaption{\label{uxclumpy_table}\textsc{UxClumpy} Fit Parameters for Rejected Models\tablenotemark{a}}
\tablehead{
  \colhead{Parameter} & \colhead{Frozen $f_{\rm scatt}$} &  \colhead{Unconstrained $f_{\rm scatt}$} \\
}
\startdata
$\Gamma$                            & 1.60$^{+0.13}_{-0.10}$ & 1.87$^{+0.13}_{-0.10}$ \\
Power law norm (10$^{-4}$)           & 5.30$^{+0.65}_{-0.16}$ & 3.26$^{+1.70}_{-1.12}$ \\
Inclination angle ($^{\circ}$)        & $<$8 & \nodata\tablenotemark{d} \\
$N_{\rm H,los}$ (10$^{24}$ cm$^{-2}$)    & 0.18$^{+0.06}_{-0.03}$ & 0.31$^{+0.07}_{-0.08}$\\
$\sigma_{\rm Torus}$ ($^{\circ}$)\tablenotemark{b} & 14.1$^{+1.7}_{-0.7}$ & $>$70 \\
CTKcover\tablenotemark{c} & $<$0.20 & $<$0.27 \\
$f_{\rm scatt}$ (10$^{-2}$)   & 10(f)  & $>$66 \\
$\chi^2$ (dof)            & 361.8 (408) & 351.7 (408) \\
\enddata
\tablenotetext{a}{We present results for two realizations of using the \textsc{UxClumpy} model to fit the X-ray spectra of 3c223: freezing the scattering fraction ($f_{\rm scatt}$) to the maximum advised limit of 10\% \citep[``Frozen $f_{\rm scatt}$'';][]{buchner2019} and allowing no limits on the scattering fraction other than a maximum value of unity (``Unconstrained $f_{\rm scatt}$'').}
\tablenotetext{b}{$\sigma_{\rm Torus}$ refers to the angular width of clouds in a clumpy torus, defining the torus scale height.}
\tablenotetext{c}{CTKcover parameterizes the covering factor of an inner Compton-thick ring of clouds ($N_{\rm H} > 10^{25}$ cm$^{-2}$) around the corona.}
\tablenotetext{d}{The best fit inclination angle was 90$^{\circ}$, but the lower limit on this value was unconstrained, with a $\Delta \chi^2$ value of 0.4 for a 90\% confidence interval for a 0$^{\circ}$ inclination angle.}
\end{deluxetable}

\clearpage
\section{Estimating Extinction-Corrected \ion{O}{3} Flux}\label{o3corr_derivation}

Dust in the AGN Narrow Line Region (NLR) causes optical emission to suffer extinction. Such extinction is wavelength dependent, with bluer wavelengths being more extinguished than redder wavelength. This effect is described by extinction laws calibrated using measured fluxes of stars in the Milky Way \citep[e.g.,][]{seaton,cardelli}, Large Magellanic Cloud \citep[e.g.,][]{howarth,fitzpatrick}, and Small Magellanic Cloud \citep[e.g.,][]{prevot,bouchet}, or from the integrated emission of gas in external galaxies \citep[e.g.][]{calzetti1994,calzetti2000}. The Balmer decrement, or the ratio of the observed H$\alpha$ to H$\beta$ flux compared to a theoretical value in the absence of extinction, provides a way to estimate and correct for the reddening of optical emission lines in the NLR.

For light embedded within a uniform dust layer, the observed flux ($F_{\rm obs}$) is related to the intrinsic flux ($F_{\rm intrinsic}$) by:
\begin{equation}
  F_{\rm obs} = F_{\rm intrinsic}\,e^{-\tau(\lambda)},
\end{equation}
where $\tau(\lambda)$ is the (wavelength-dependent) optical depth. The difference in optical depth between the H$\alpha$ and H$\beta$ emission lines ($\tau_{\rm H\beta}  - \tau_{\rm H\alpha}$) is then:
\begin{equation}
 \tau_{\rm H\beta} - \tau_{\rm H\alpha} = {\rm ln}\left(\frac{F_{\rm H\alpha,obs}/F_{\rm H\beta,obs}}{F_{\rm H\alpha,intrinsic}/F_{\rm H\beta,intrinsic}}\right).
\end{equation}
We take $F_{\rm H\alpha,intrinsic}/F_{\rm H\beta,intrinsic}$ to be 3.1, consistent with the temperatures and densities observed within the AGN NLR gas \citep{osterbrock2006}.

We then use the \citet{cardelli} extinction law ($k(\lambda)$) to estimate the level of extinction in the AGN NLR given this Balmer decrement. The \citet{cardelli} extinction law has the form:
\begin{equation}
  k(\lambda) = A(\lambda)/A(V) = a(x)\, + \, b(x)/R_V,
\end{equation}
where $A(\lambda)$ is the extinction (in magnitudes) at any wavelength $\lambda$, $A(V)$ is the reference extinction in the $V$ band, $R_V \equiv A(V)/E(B-V)$ (where $E(B-V)$ is the color excess due to reddening), and $a(x)$ and $b(x)$ are polynomials parameterizing the extinction law in units of $x = 1/\lambda$ ($\mu$m$^{-1}$) \citep[see Equations 3a and 3b in][for the optical-near-infrared extinction law]{cardelli}. In the \citet{cardelli} model, $R_V = 3.1$. Since $A(\lambda)$ is measured in units of magnitude, i.e.,:
\begin{equation}
  A(\lambda) = -2.5\,{\rm log}\left(\frac{F_{\rm obs}}{F_{\rm intrinsic}}\right),
\end{equation}
the relationship between extinction in magnitudes and optical depth is:
\begin{equation}
  A(\lambda) = 1.086\,\tau
\end{equation}

Combining the above equations, $A(V)$ can be cast in terms of the extinction law and Balmer optical depth with the following:
\begin{equation}
  A(V) = A(\lambda_{\rm Balmer\, decrement})/k(\lambda_{\rm Balmer\, decrement}) = \frac{1.086\,(\tau_{\rm H\beta} - \tau_{\rm H\alpha})}{k(\rm H\beta) - k(\rm H\alpha)},
\end{equation}
where $k(\rm H\beta)$ and $k(\rm H\alpha)$ are the values of the \citet{cardelli} extinction law at the wavelengths of H$\beta$ and H$\alpha$, respectively, giving:
\begin{equation}
  A(V) = \frac{1.086\,(\tau_{\rm H\beta} - \tau_{\rm H\alpha})}{1.164\, - 0.818} = 3.14 \,(\tau_{\rm H\beta} - \tau_{\rm H\alpha}) = 3.14 \, {\rm ln}\left(\frac{F_{\rm H\alpha,obs}/F_{\rm H\beta,obs}}{3.1}\right).
\end{equation}

Finally, we can measure the optical extinction at any wavelength with:
\begin{equation}
  A(\lambda) = k(\lambda)A(V) = k(\lambda) \times 3.14\,{\rm ln}\left(\frac{F_{\rm H\alpha,obs}/F_{\rm H\beta,obs}}{3.1}\right).
\end{equation}

With $k$(5007\AA) = 1.12, the optical depth to the [\ion{O}{3}] line is:
\begin{equation}
  \tau_{\rm 5007 \AA} = A(5007{\rm \AA})/1.086 = 3.2\,{\rm ln}\left(\frac{F_{\rm H\alpha,obs}/F_{\rm H\beta,obs}}{3.1}\right),
\end{equation}
which provides the relation quoted in the main text of:
\begin{equation}
  F_{\rm [O\,III],intrinsic} = F_{\rm [O\,III],obs}\left(\frac{F_{\rm H\alpha,obs}/F_{\rm H\beta,obs}}{3.1}\right)^{3.2}.
\end{equation}

\end{document}